\documentclass[11pt]{article}
\usepackage{amsfonts}
\usepackage{mathrsfs}
\usepackage[latin1]{inputenc}
\usepackage{float}
\usepackage{amsmath,amssymb}
\usepackage{latexsym}
\usepackage{epstopdf}
\usepackage{geometry}  
\usepackage[active]{srcltx}
\usepackage{graphicx}
\usepackage{epstopdf}
\usepackage{booktabs}
\usepackage{multirow}
\usepackage{siunitx}

\usepackage[
bookmarks=true,         
bookmarksnumbered=true, 
colorlinks=true, pdfstartview=FitV, linkcolor=blue, citecolor=blue,
urlcolor=blue]{hyperref}

\providecommand{\U}[1]{\protect\rule{.1in}{.1in}}
\newtheorem {theorem}{Theorem}[section]

\newtheorem{definition}{Definition}[section]

\newtheorem{remark}{Remark}[section]

\newcommand{\E}{\mathbb{E}}

\newcommand{\bi}[1]{\mbox{\boldmath{$ #1 $}}}

\begin{document}
\title{A New Gini Correlation between Quantitative and Qualitative Variables}
\author{Xin Dang\textsuperscript{a}\thanks{CONTACT Xin Dang. Email: xdang@olemiss.edu}, Dao Nguyen\textsuperscript{a},Yixin Chen\textsuperscript{b} and Junyin Zhang\textsuperscript{c}}
\date{%
    \textsuperscript{a}Department of Mathematics, University of Mississippi, University, MS 38677, USA\\%
    \textsuperscript{b}Department of Computer and Information Science, University of Mississippi, University, MS 38677, USA\\%
     \textsuperscript{c}Department of Mathematics, Taiyuan University of Technology, Taiyuan, 030024, P. R. China\\[1ex] %
  \today}
\maketitle
\begin{abstract}
We propose a new Gini correlation to measure dependence between a categorical and numerical variables.  Analogous to Pearson $R^2$ in ANOVA model, the Gini correlation is interpreted as the ratio of the between-group variation and the total variation, but it characterizes independence (zero Gini correlation mutually implies independence). Closely related to the distance correlation, the Gini correlation is of  simple formulation by considering the nature of categorical variable. As a result, the proposed Gini correlation has a lower computational cost than the distance correlation and is more straightforward to perform inference. Simulation and real applications are conducted to demonstrate the advantages.  
\noindent  
\vskip.2cm 
\noindent {\bf Keywords:}
\noindent Distance correlation, Energy distance, Gini mean difference, Gini correlation

\end{abstract}


\section{Introduction}

Measuring strength of association or dependence between two variables or two sets of variables is of vital importance in many research fields.  
Various correlation notions have been developed and studied \cite{Kendall90, Mari01}. The widely-used Pearson product correlation measures the linear relationship. Rank based or copula based correlations such as Spearman's $\rho$ \cite{Spearman04} and Kendall's $\tau$ \cite{Kendall38} explore monotonic relationships. Gini correlation \cite{Schechtman87, Schechtman03} is based on the covariance of one variable and rank of the other.  A symmetric version of Gini correlation is proposed by Sang, Dang and Sang (2016) \cite{Sang16}.  Other robust correlation measures are surveyed in \cite{Devlin75, Shevlyakov11} and explored in detail in \cite{Shevlyakov16}. The distance correlation proposed by Sz\'{e}kely and Riozzo (2009) \cite{Szekely09} characterizes dependence for multivariate data. Those correlations, however, are only defined for numerical and/or ordinal variables. They can not be directly applied to a categorical variable.   

If both variables are nominal, Cram\'{e}r's $V$ \cite{Cramer46} and Tschuprow's $T$ \cite{Tschuprow39} based on $\chi^2$ test statistic can be used to measure their association. Theoretically based on information theory, mutual information is popular due to its easy computation for two discrete variables. However, mutual information correlation \cite{Ross14, Gao17} loses the computational attractiveness for measuring dependence between categorical and numerical variables, especially when the numerical variable is in high dimension. 

For this case, two approaches are typically used for defining association measures. The first one treats the continuous numerical variable $X$ as the response variable and the categorical variable $Y$ as the predictor. Pearson $R^2$ of the analysis of variance (ANOVA) or $\eta^2$ of MANOVA is then a measure of correlation between them. The second approach considers $Y$ being the response and $X$ as the explanatory variable(s).  A pseudo-$R^2$ of the logistic or other generalized regression model serves a measure of correlation \cite{Tjur09}. If $X$ and $Y$ are independent, those correlation parameters are zero. However, the converse is not true in general. Those correlations do not characterize independence. In this paper, we propose a new Gini correlation (denoted as $\rho_g$) for measuring dependence between categorical and numerical variables.  

The contributions of this paper are as follows. 
\begin{itemize}
\item A  new  dependence measure between categorical and numerical variables. The proposed Gini correlation characterizes independence: zero correlation mutually implies independence. It also has a nice interpretation as the ratio of between Gini variation and the total variation. 
\item Limiting distributions of sample Gini correlation obtained under independence and dependence cases.  
\item Extension of the distance correlation for dependence measure between categorical and numerical variables.
\item Comparison of Gini correlation and the distance correlation. Comparing with the distance correlation, Gini correlation has a simpler form, leading a simple computation and easy inference. 
\end{itemize}

The remainder of the paper is organized as follows. Section \ref{sec:mot} motivates a dependence measure between one-dimensional numerical variable and a categorical variable. The connection to Gini mean difference leads to a natural generalization and nice interpretation. The properties of the generalized Gini correlation are studied in Section \ref{sec:ggc}. The relationship to the distance correlation is treated in Section \ref{sec:dc} and three examples are given in Section \ref{sec:example}. Section \ref{sec:inference} is devoted to inferences of the Gini correlation. Asymptotic behavior of the sample Gini correlation is explored. In Section \ref{sec:experiment}, we conduct experimental studies by simulation and real data applications to demonstrate advantages of the Gini correlation over the distance correlation. We conclude and discuss future works in Section \ref{sec:conclusion}. Proofs and detailed derivations of Remarks and Example results are provided in the Online Supplement.  

\section{Motivation}
\label{sec:mot}
\subsection{Proposed correlation}
We consider to measure association between a numerical variable $ X$ in $\mathbb{R}$ and a categorical variable $Y$.  Suppose that $Y$ takes values $L_1, ...,L_K$. Assume the categorical distribution $P_Y$ of $Y$ is $P(Y = L_k) = p_k>0$ and the conditional distribution of $ X$ given $Y=L_k$ is $F_k$. Then the joint distribution of X and Y is $P(X\leq x, Y=k)= p_kF_k(x)$. When the conditional distribution of $X$ given $Y$ is the same as the marginal distribution of $X$, $X$ and $Y$ are independent. In that case, we say there is no correlation between them. However, when they are dependent, i.e $F(x) \neq F_k( x)$ for some $k$, we would like to measure this dependence.  Intuitively, the larger the difference between the marginal distribution and conditional distribution is, the stronger association should be. With that consideration, a natural correlation measure shall be proportional to 
\begin{equation}\label{distD}
D: = \E \int_{\mathbb{R}} (F(x|Y)-F(x))^2\, d x = \sum_{k=1}^K p_k \int_{\mathbb{R}} (F_k( x) -F(x))^2\, d x, 
\end{equation}
the expectation of the integrated squared difference between conditional and marginal distribution functions, if $D$ is finite.  In other words, the correlation is proportional to the $L_2$ distance of marginal and conditional distributions.  

Clearly, the corresponding correlation is non-negative, just like Pearson $R^2$ type of correlations. It, however, has an advantage that the correlation is zero if and only if $X$ and $Y$ are independent, while for Pearson $R^2$ type of correlation, zero does not mutually imply independence. 

Next,  we need to find the standardization term so that the corresponding correlation has a range of $[0,1]$, a desired property for a dependence measure specified in  \cite{Renyi59}. In other words, under some condition of $F$, we would like to obtain $\max D$ among all $F_k$ and $p_k$, which can be formulated to solve the following optimization problem. 
\begin{align} \label{maxD}
&\max_{F_k, p_k} D = \max_{F_k, p_k}  \sum_{k=1}^K p_k \int_{\mathbb{R}} (F_k( x)-F(x))^2\, d x,\\
&\mbox{subject to } p_k > 0, \sum_{k=1}^K p_k =1,  \sum_{k=1}^K p_kF_k( x) = F(x)\nonumber\\ 
& \mbox{and }F_k(x) \mbox{ is a distribution function for }k=1,...,K. \nonumber 
\end{align}
Note that $\sum_{k=1}^K p_k(F_k( x)-F( x))^2 =\sum_{k=1}^K p_k F_k^2( x) -F^2( x)  \geq 0$ for any $ x$. Since $F_k( x)$ is a cumulative distribution function, we have
$$D = \int_{\mathbb{R}} \sum_{k=1}^K p_k F_k^2( x) -F^2( x) d  x \leq \int_{\mathbb{R}} F( x) - F^2( x) d x. $$
The equality holds if and only if $F_k$ is a single point mass distribution.  In that case, $F$ is a discrete distribution with at most $K$ distinct values almost surely.  Assuming that $0< \int_{\mathbb{R}} F(x) - F^2( x) \,d x <\infty$, we propose the correlation between $ X$ and $Y$ as
\begin{equation} \label{gc}
\rho(X,Y) = \frac{\sum_{k=1}^K p_k \int_{\mathbb{R}} (F_k( x) -F( x))^2\, d  x}{\int_{\mathbb{R}} F( x) - F^2( x)\, d x}. 
\end{equation}
From the discussion above, we have the following immediate results. 
\begin{enumerate}
\item $0\leq \rho( X,Y) \leq 1$.
\item $\rho( X,Y) = 0$ if and only if $X$ and $Y$ are independent.
\item $\rho(X,Y)=1 $ if and only if $F_k$ is a single point mass distribution.  
\end{enumerate}

Assumption $\int_{\mathbb{R}} F(x) - F^2( x) \,d x > 0$ implies that $F$ is not a point mass distribution and hence $X$ is non-degenerate. Assumption $\int_{\mathbb{R}} F(x) - F^2( x) \,d x < \infty$ means $\E |X|< \infty$, which we will see in the next subsection.  Further, $\rho( X,Y)$ can be written as
\begin{equation}\label{gc2}
\rho( X, Y)= 1-\frac{2\sum_{k=1}^K p_k  \int_{\mathbb{R}}  F_k(x)-F_k^2( x) \,d  x}{2\int_{\mathbb{R}} F( x) - F^2(x)\; d x}.
\end{equation}
This formulation provides a Gini mean difference representation of the proposed correlation.

\subsection{Gini distance representation}
Gini mean difference (GMD) was introduced as an alternative measure of variability to the usual standard deviation (\cite{Gini14}, \cite{David68}, \cite{Yitzhaki13}). Let $X$ and $X^\prime$ be independent random variables from a distribution $F$ with finite first moment in $\mathbb{R}$.  The GMD of $F$ is
\begin{equation}\label{eqn:gmd}
\Delta=\Delta(X)= \Delta(F)=\E |X-X^\prime|,
\end{equation}
the expected distance between two independent random variables. Dorfman (1979) \cite{Dorfman79} proved that for non-negative random variables,
\begin{equation}\label{deltaF}
\Delta= 2 \int F(x) (1-F(x)) \,d x.
\end{equation}
The proof can be easily extended to any random variable with $\E|X|<\infty$ \cite{Yitzhaki13}. Note that (\ref{deltaF}) also holds for discrete random variables. Hence,  we can write the correlation of (\ref{gc2}) as
\begin{equation}\label{gc3}
\rho(X,Y) = 1- \frac{\sum_{k=1}^K p_k\Delta_k}{\Delta} = \frac{\Delta - \sum_{k=1}^K p_k\Delta_k}{\Delta},
\end{equation}
where $\Delta$ is the Gini mean difference (GMD) of $F$ and $\Delta_k$ is the GMD of $F_k$.  We call it the Gini correlation and denote as $\rho_g(X,Y)$ or $gCor(X,Y)$. 

The representation of (\ref{gc3}) allows another interpretation. Consider that $\sum_{k=1}^K p_k\Delta_k$, the weighted average of Gini mean differences,  is a measure of within-group variation and $\Delta - \sum_{k=1}^K p_k\Delta_k$ is the corresponding between group variation. The proposed correlation is the ratio of the between-group Gini variation and the total Gini variation, analogue to the Pearson $R^2$ correlation in ANOVA (Analysis of Variance). The squared Pearson correlation is defined to be the ratio of between variance and the total variance. Denote $\mu, \sigma^2, \mu_k$, and $\sigma_k^2$ as the mean and variance of $F$ and $F_k$, respectively.  The variance of $X$ can be partitioned to the within variation and the between variation as below,  
$$\sigma^2 = Var(X)= \E[\E X^2|Y] -(\E[\E X|Y])^2 = \sum_{k=1}^K p_k(\sigma_k^2+\mu_k^2) -\mu^2 = \sum_{k=1}^K p_k \sigma_k^2 + (\sum_{k=1}^K p_k \mu_k^2 -\mu^2). $$
And Pearson $R^2$ correlation, denoted as $\rho_p^2(X,Y)$,  is 
$$\rho_p^2 (X,Y) = 1- \frac{\sum_{k=1}^K p_k \sigma_k^2}{\sigma^2} = \frac{ \sum_{k=1}^K p_k \mu_k^2 -\mu^2}{\sigma^2}.$$
Let $(X,X^{\prime})$, $(X_k, X_k^{\prime})$, $(X_l, X_l^{\prime})$ be  independent pair variables independently from $F$, $F_k$ and $F_l$, respectively. It is easy to derive that 
\begin{equation} \label{gmd}
\Delta = \E|X-X^{\prime}| =\E \E(|X-X^{\prime}||Y, Y^\prime)=\sum_{k=1}^Kp_k^2 \Delta_k +2\sum_{1\leq k<l\leq K} p_kp_l\Delta_{kl}, 
\end{equation}
where $ \Delta_k = \E |X_k-X_k^{\prime}|$ and $ \Delta_{kl }= \E |X_k-X_l|$. Then the between Gini variation, denoted as the Gini distance covariance between $X$ and $Y$, is 
\begin{equation}\label{gcov}
\mbox{gCov} (X,Y) = \Delta- \sum_{k=1}^Kp_k \Delta_k = 2\sum_{1\leq k<l\leq K} p_kp_l\Delta_{kl} -\sum_{k=1}^Kp_k(1-p_k)\Delta_k, 
\end{equation}
and 
the Gini distance correlation between $X$ and $Y$ is 
\begin{equation}\label{rg1}
\mbox{gCor}(X,Y)= \rho_g(X,Y) = \frac{gCov(X,Y)}{\Delta(X)}. 
\end{equation}

The total Gini variation is partitioned to the within and the between Gini variation.   Frick et al. (2006) \cite{Frick06} consider another decomposition of the Gini variation, which is represented by four components, i.e, within Gini variation,  between Gini variation among group means and two effects of overlapping among groups. Although the extra terms provide some insights of the extent of group intertwining, their decomposition is complicated.  Not only our representation of the total Gini variation is simple and easy to interpret, but also it is natural to extend to the multivariate case. 

\section{Proposed Gini Correlation}
\label{sec:ggc}
\subsection{Generalized Gini Correlation}
There are two multivariate generalizations for the Gini mean difference. One is the Gini covariance matrix proposed by Dang et al. (2019) \cite{Dang16}. Along this line, one may extend the Gini correlation based on an analog of Wilk's lambda or Hotelling-Lawley trace in MANOVA. That leaves for future work. Here we explore another generalization defined in \cite{Koshevoy97}.  That is, the Gini mean difference of a distribution $F$ in $\mathbb{R}^d$ is  
$$ \Delta =\E \|\bi X -\bi X ^\prime\|, $$
or even more generally for some $\alpha$,
\begin{equation}  \label{ggmd}
 \Delta(\alpha) = \E \|\bi X-\bi X^\prime\|^{\alpha}, 
 \end{equation} 
where $\| \bi x \|$ is the Euclidean norm of $\bi x$. With this generalized multivariate Gini mean difference (\ref{ggmd}), we can define the Gini correlation in (\ref{gc2}) as follows.
\begin{definition} 
For a non-degenerate random vector $\bi X$ in $\mathbb{R}^d$ and a categorical variable $Y$, if  $\E\|\bi X\|^\alpha <\infty$ for $\alpha \in (0, 2)$, the Gini correlation of $\bi X$ and $Y$ is defined as
\begin{equation} \label{mgc}
\rho_g(\bi X, Y; \alpha) =1-\frac{\sum_{k=1}^K p_k\Delta_k(\alpha)}{\Delta(\alpha)}= \frac{\Delta(\alpha) -\sum_{k=1}^K p_k\Delta_k(\alpha)}{\Delta(\alpha)}, 
\end{equation}
where $\Delta_k(\alpha)$ and $\Delta(\alpha)$ are the generalized Gini differences of distribution $F_k$ and $F$, respectively. 
\end{definition}

\begin{remark}\label{alpha}
Note that a small $\alpha>0$ provides a weak assumption of  $\E \|\bi X\|^\alpha< \infty $ on distributions, which allows applications of the Gini correlation to heavy-tailed distributions. 
\end{remark}

\begin{remark}
If $\alpha =2$ and $d=1$,  $\rho_g(X,Y; 2) = \rho_p^2(X,Y)$ because of the fact that $\Delta(2)=\E |X-X^\prime|^2 = 2 Var(X)$.  The requirement of  $\alpha \in (0,2)$ is for desired properties of the Gini correlation. 
\end{remark}

The next theorem states the properties of the proposed Gini correlation. 
\begin{theorem} \label{mgcor}
For a categorical variable $Y$ and a continuous random vector $\bi X$ in $\mathbb{R}^d$ with $\E \|\bi X\|^{\alpha} <\infty$ for $0<\alpha<2$, $\rho_g(\bi X, Y;\alpha)$ has following properties. 
\begin{enumerate}
    \item $0\leq \rho_g(\bi X,Y; \alpha) \leq 1$.
    \item $\rho_g(\bi X,Y; \alpha) =0 $ if and only if $\bi X$ and Y are independent.
    \item $\rho_g(\bi X,Y; \alpha) =1 $ if and only if $F_k$ is a single point mass distribution for $k=1,...,K$. 
    \item $\rho_g(a O\bi X+\bi b, Y; \alpha)= \rho_g(\bi X, Y;\alpha) $ for any orthonormal matrix $O$ $(O^T = O^{-1})$, nonzero constant $a$ and vector $\bi b$. 
\end{enumerate}
\end{theorem}

{\bf Proof.} First of all, $\Delta_k(\alpha)\geq 0$, so we have $\rho_g(\bi X,Y;\alpha) \leq 1$.  It is obvious that $\rho_g(\bi X,Y;\alpha) = 1$ if and only if  $\Delta_k(\alpha)=0$ for each $k$, which mutually implies  that $F_k$ is a singleton distribution.  Orthogonal invariance of the Property (4) is a result from the Euclidean distance used in the Gini correlation.  The proof for the remaining part has two steps.   In Step 1,  we can write 
\begin{equation}\label{gce}
\mbox{gCov}(\bi X,Y;\alpha) = \Delta(\alpha) -\sum_{k=1}^Kp_k\Delta_k(\alpha) = \sum_{k=1}^K p_k T(\bi X_k, \bi X; \alpha)
\end{equation}
where  
$T(\bi X_k, \bi X;\alpha)= 2 \E\|\bi X_k -\bi X\|^{\alpha} -\E\|\bi X_k -\bi X_k^\prime\|^{\alpha} -\E \|\bi X-\bi X^\prime\|^{\alpha}.$  
This is because 
\begin{eqnarray}
 \sum_{k=1}^{K} p_k T(\bi X_k, \bi X; \alpha) &=& \sum_{k=1}^K p_k (2 p_k \Delta_k(\alpha) +2 \sum_{l \neq k} p_l \Delta_{kl}(\alpha) -\Delta_k(\alpha) - \Delta(\alpha))   \nonumber\\
  &=&\sum_{k=1}^K (2p_k^2-p_k)\Delta_k(\alpha)+ 2 \sum_{k=1}^K\sum_{l \neq k} p_kp_l\Delta_{kl}(\alpha)  -\Delta(\alpha)\nonumber \\
& =& 2\sum_{1\leq k<l\leq K} p_kp_l\Delta_{kl}(\alpha) -\sum_{k=1}^Kp_k(1-p_k)\Delta_k(\alpha)    \label{equ14}\\ 
&=&\Delta(\alpha) -\sum_{k=1}^Kp_k \Delta_k(\alpha).\nonumber
\end{eqnarray}
The third equality (\ref{equ14}) is obtained by plugging in (\ref{gmd}) and the last equality is due to (\ref{gcov}).   In Step 2, one recognizes that $T(\bi X_k,\bi X,\alpha)$ is the energy distance between $\bi X$ and $\bi X_k$ defined in Sz\'{e}kely and Rizzo (2013, 2017). Applying the Proposition 2 of Sz\'{e}kely and Rizzo (2013), for $0<\alpha<2$, we have 
\begin{align} \label{charac}
&T(\bi X_k, \bi X; \alpha)= c(d,\alpha)  \int_{\mathbb{R}^d} \frac{|\psi_k(\bi t) -\psi(\bi t)|^2}{\|\bi t\|^{d+\alpha}} d\bi t,
\end{align}
where $\psi_k$ and $\psi$ are the characteristic functions of $\bi X_k$ and $\bi X$, respectively, and $c(d,\alpha)$ is a constant only depending on $d$ and $\alpha$, i.e.,
$$c(d,\alpha) = \frac{\alpha 2^{\alpha} \Gamma((d+\alpha)/2)}{2\pi^{d/2} \Gamma(1-\alpha/2)}.$$
Results of \eqref{gce} and \eqref{charac}  show that for all $0<\alpha<2$, we have $\mbox{gCov}(\bi X, Y; \alpha)\geq 0$ and hence $\rho_g(\bi X, Y, \alpha) \geq 0$ with equality to zero if and only if $\bi X$ and $\bi X_k$ are identically distributed for all $k=1,...,K$.  
\hfill{$\square$}

\medskip
Below we provide a couple of remarks and their proofs are given in the Online Supplement.
\begin{remark}\label{rem:2-3}
$T(\bi X_k, \bi X; \alpha)$ is the energy distance of $\bi X_k$ and $\bi X$, which is the weighted $L_2$ distance of characteristic functions of $\bi X_k$ and $\bi X$. For $d=1$, $T(X_k,X; 1)$ is also the $L_2$ distance of the distribution function $F_k$ and $F$ multiplying a constant.  However, such a relationship does not hold for $d>1$.  
\end{remark}
\begin{remark}\label{rem:2-4}
The Gini covariance of $\bi X$ and $Y$ is the weighted average of energy distance between $\bi X_k$ and $\bi X$.  It  is also a linear combination of energy distances between $\bi X_k$ and $\bi X_l$. That is, $\mbox{gCov}(\bi X,Y; \alpha) = \sum_{k=1}^K p_k T(\bi X_k, \bi X; \alpha)= \sum_{1 \leq k<l\leq K} p_kp_l T(\bi X_k, \bi X_l; \alpha) $.  
\end{remark}

Particularly for $K=2$,  the between variation  $\mbox{gCov}(\bi X, Y;\alpha) = \Delta(\alpha)- p_1\Delta_1(\alpha)-p_2\Delta_2(\alpha)$,  is simplified to be 
\begin{align*}
& p_1 T(\bi X_1, \bi X;\alpha)+p_2 T(\bi X_2, \bi X;\alpha)
= p_1p_2 T(\bi X_1, \bi X_2;\alpha) 
\end{align*}
which is  proportional  to $T(\bi X_1, \bi X_2; \alpha)$, the energy distance used in \cite{Szekely13, Szekely17}.  Sz{\'e}kely and Rizzo \cite{Szekely04} considered a special case of the energy distance of $\alpha=1$ and proposed a test for the equality of two distributions $F_1$ and $F_2$, which is also studied in \cite{Baringhaus04}. The test is equivalent to test $\rho_g(\bi X,Y; \alpha)=0$.  The test of $\rho_g(\bi X,Y; \alpha)=0$ is also used for the $K$-sample problem. In that case, it is equivalent to the test of DISCO (DIStance COmponent) analysis in \cite{Rizzo10}. The test statistic in DISCO takes the ratio of the between and the within group Gini variations for the $K$-sample problem. Testing $\rho_g(\bi X, Y;\alpha) = 0$ is equivalent to the one-way DISCO analysis.  What we contribute in the dependence test is that our test is able to provide power analysis for a particular alternative if it is specified as $\rho_g (\bi X, Y; \alpha)= \rho_0$ where $\rho_0> 0$.

\subsection{Connection to the Distance Correlation}
\label{sec:dc}
The proposed Gini correlation is closely related to but different from the distance correlation studied by Sz\'{e}kely, Rizzo and Bakirov (2007) \cite{Szekely07},  Sz\'{e}kely and Rizzo (2009) \cite{Szekely09}. Their distance correlation considers correlation between two sets of continuous random variables. Later the distance covariance and distance correlation are extended from Euclidean space to general metric spaces by Lyons (2013) \cite{Lyons13}. Based on that idea, we define the discrete metric 
$$d(y, y^\prime) =|y-y^\prime|:=  I(y\neq y^\prime),$$
where $I (\cdot)$ is the indicator function. Equipped with this set difference metric on the support of $Y$ and Euclidean distance on the support of $\bi X$, the corresponding distance covariance and distance correlation for numerical $\bi X$ and categorical $Y$ variables are as follows. 
\begin{align}
&\mbox{ dCov}(\bi X,Y; \alpha)=  c(d,
\alpha)\sum_{k=1}^K \int \frac{(p_k\psi_k(\bi t) -p_k\psi(\bi t))^2}{\|\bi t\|^{d+\alpha}} d\bi t, \label{dcov}\\
& \mbox{ dCov}(\bi X,\bi X;\alpha) =c(d,
\alpha)^2 \int \frac{(\psi(\bi t+\bi s) -\psi(\bi t)\psi(\bi s))^2}{\|\bi t\|^{d+\alpha}\|\bi s\|^{d+\alpha}} d\bi t d\bi s,  \nonumber \\
& \mbox{ dCov}(Y,Y)= \sum_{k=1}^K p_k^2-2\sum_{k=1}^Kp_k^3+(\sum_{k=1}^K p_k^2)^2, \label{dcovyy}\\
&\rho_d(\mathbf{X}, Y; \alpha):= \mbox{ dCor}(\bi X, Y; \alpha) = \frac{\mbox{ dCov}(\bi X, Y, \alpha)}{ \sqrt{\mbox{ dCov}(\bi X,\bi X;\alpha)} \sqrt{\mbox{ dCov}(Y,Y)}}. \nonumber
\end{align}

\begin{remark} \label{identity}
As expected,  $\mbox{dCov}(\bi X, Y;\alpha) = \E |\bi X-\bi X^\prime|^\alpha|Y-Y^\prime|^\alpha+\E |\bi X-\bi X^\prime|^\alpha\E|Y-Y^\prime|^\alpha-2\E |\bi X-\bi X^\prime|^\alpha|Y-Y^{\prime\prime}|^\alpha$, where $(\bi X, Y), (\bi X^\prime, Y^\prime), (\bi X^{\prime\prime}, Y^{\prime\prime})$ are i.i.d.  
\end{remark}
 The proofs of this identity in Remark \ref{identity}, (\ref{dcovyy}) and the following Remark \ref{rem:2class} are given in the Online Supplement. 
Comparing (\ref{dcov}) with (\ref{gce}) and (\ref{charac}), it is easy to make the following conclusions.
\begin{remark}\label{dcxye}
$dCov(\bi X, Y;\alpha) = \sum_{k=1}^K p_k^2 T(\bi X_k, \bi X;\alpha)$. 
\end{remark}

    \begin{remark}
    $gCov(\bi X, Y;\alpha)\geq dCov(\bi X, Y;\alpha)$. They are equal if and only if $\bi X$ and $Y$ are independent with both being zero. 
    \end{remark}
    \begin{remark}\label{balance}
     When $p_1=p_2=...=p_K =\frac{1}{K}$,  $gCov(\bi X, Y; \alpha)=K dCov(\bi X, Y;\alpha)$.
     \end{remark}
     \begin{remark}\label{rem:2class}
     For $K=2$, $gCov(\bi X,Y; \alpha) =\frac{1}{ 2 p_1p_2} dCov(\bi X,Y;\alpha)$ and $dCov(Y,Y) = 4p_1^2p_2^2$, i.e., $gCov(\bi X,Y; \alpha) = dCov(\bi X,Y;\alpha)/\sqrt{dCov(Y,Y)}. $
     \end{remark}

\begin{remark} \label{dgd}
For the case of  $d=1$, $dCov(X,X; 1)$ is studied in \cite{Edelmann17} and  
\begin{equation}\label{dcov1}
    dCov(X,Y;1) =2  \sum_{k=1}^K \int (p_kF_k(x)-p_kF(x))^2\,dx.
\end{equation}
\end{remark}

Comparison of (\ref{dcov1}) and (\ref{distD}) explains the difference of our Gini approach and distance correlation approach in the one dimensional case. The distance covariance of $X$ and $Y$ is based on squared difference of the joint distribution $p_k F_k( x)$ and the product of the marginal distributions $p_kF( x)$, while the Gini one is based on the squared difference between the conditional distribution $F_k( x)$ and the marginal distribution $F(x)$. Our Gini dependence measure considers the categorical nature of $Y$ and has a simpler formulation than the distance correlation, leading a simpler inference and computation.  

Before we discuss their computation and inference, let us first demonstrate the Gini correlation and distance correlation in several examples. 

\subsection{Examples} 
\label{sec:example}
 Three examples for $K=2$, $d=1$ and $\alpha=1$ are provided. Denote $p_1$ as $p$. The detailed derivations and proofs for the example results are provided in the Online Supplement. \\
  \begin{figure}{h} 
 \center
\begin{tabular}{cc}
\includegraphics[width=6cm,height=6cm]{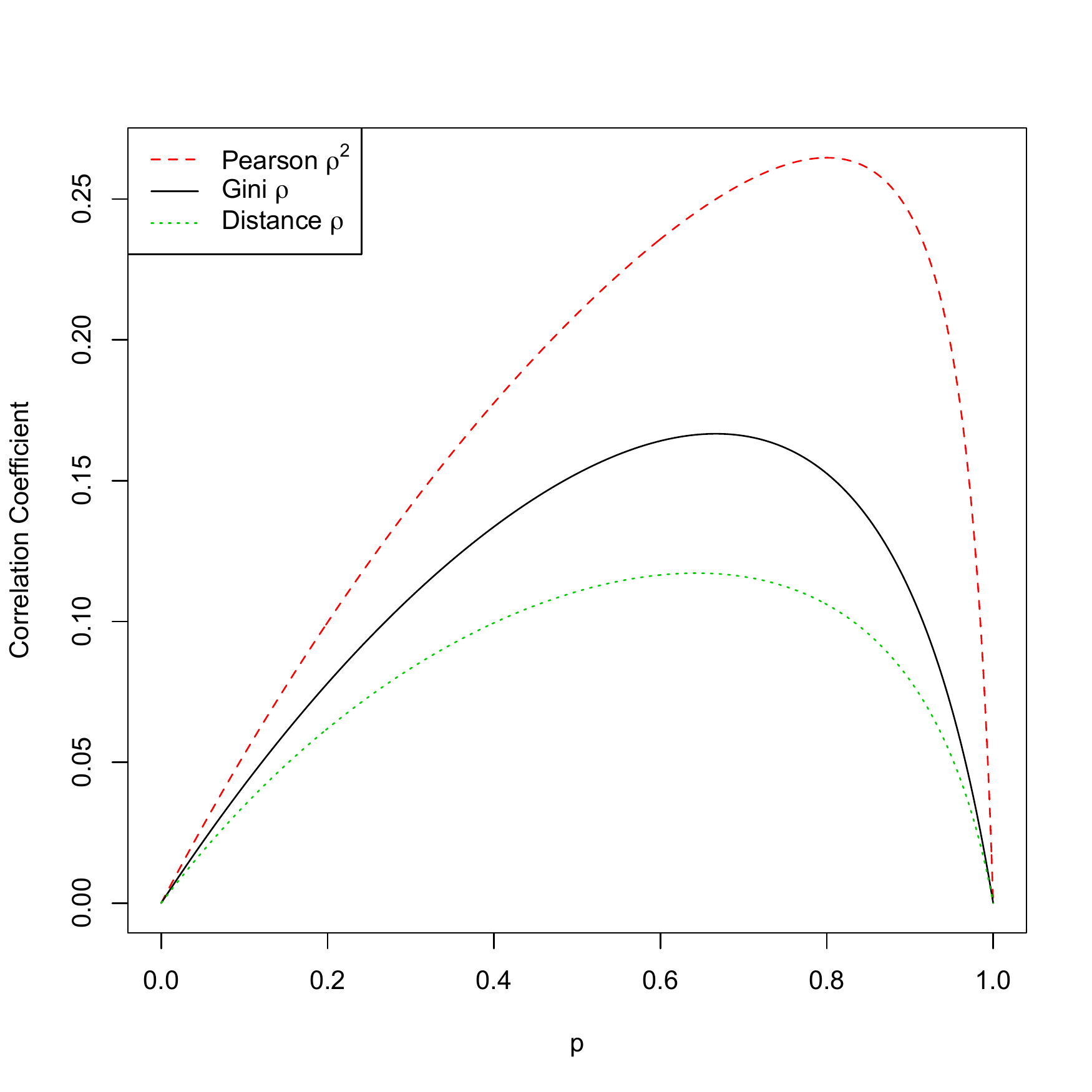}&
\includegraphics[width=6cm,height=6cm]{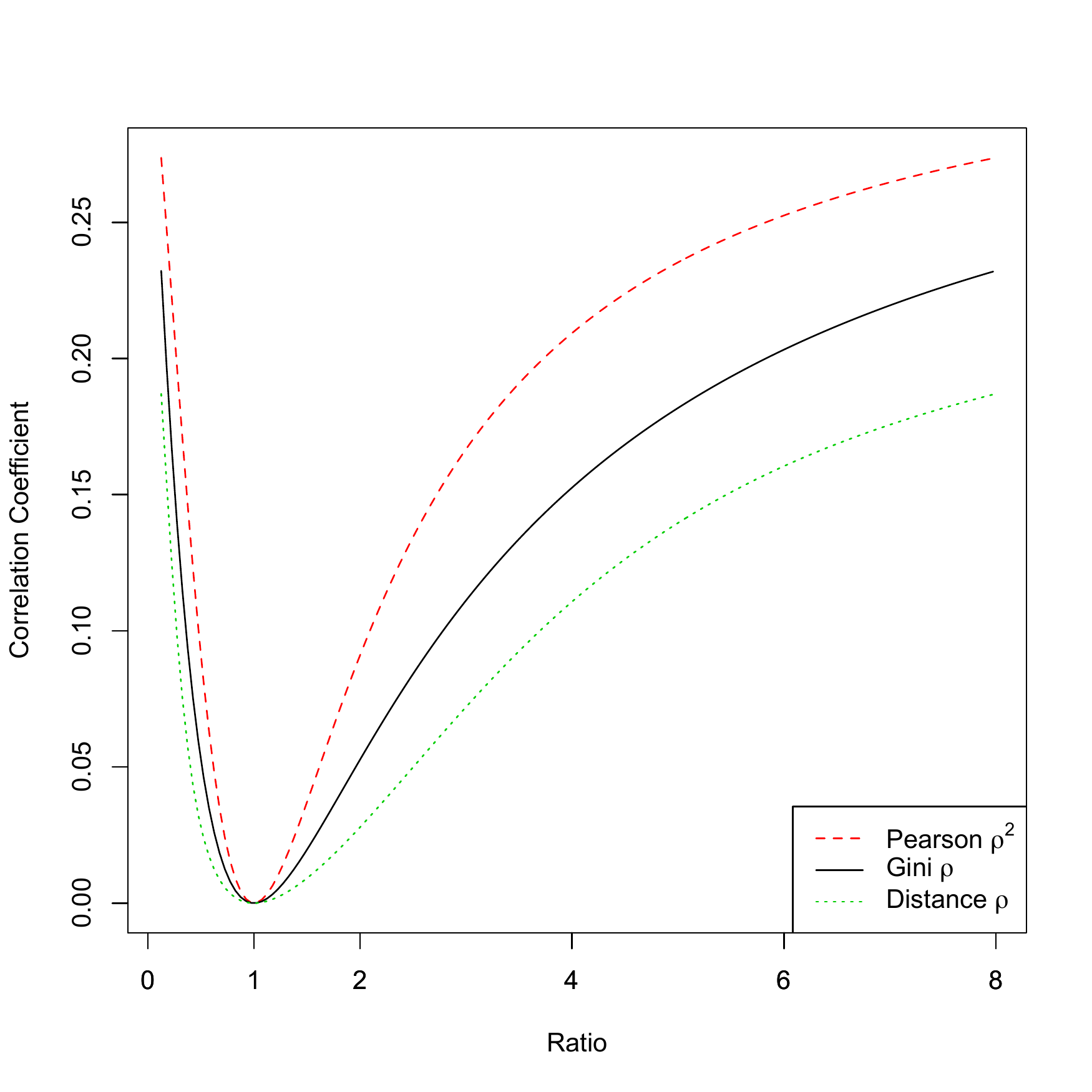}
\vspace{-0.05in} \\ 
(a)&(b)
\end{tabular}
\caption{(a) Correlation coefficients vs $p$ in the mixture exponential distribution with $\theta=1$ and $\beta = 4$; (b) Correlation coefficients vs $r = \beta/\theta$ in the mixture exponential distribution with $p=0.5$. \label{fig:exp}}
\end{figure}

\noindent {\bf Example 1.}  Let $F_1 = \mbox{Exp}(\theta)$ and $F_2= \mbox{Exp}(\beta)$. 
We have 
\begin{align*}
 &  \mu_1= \sigma_1 = \Delta_1  = \theta, \mu_2= \sigma_2 = \Delta_2 = \beta, \Delta_{12} = \frac{\theta^2+\beta^2}{\theta+\beta},\\
  & \mbox{dCov}(X,X) 
  = 2p^2\theta^2+2(1-p)^2\beta^2+(p^2\theta+(1-p)^2\beta)^2 -\frac{8}{3}p^3\theta^2-\frac{8}{3}(1-p)^3\beta^2
+\frac{16p(1-p)\theta^2\beta^2}{(\theta+\beta)^2}\\
&+\frac{32p^2(1-p)^2\theta^2\beta^2}{(\theta+\beta)^2}+\frac{8p^3(1-p)\theta^2\beta}{\theta+\beta}+\frac{8p(1-p)^3\theta\beta^2}{\theta+\beta}
-\frac{8p(1-p)^2\theta\beta^2(5\theta+\beta)}{(2\theta+\beta)(\theta+\beta)}\\
&-\frac{8p^2(1-p)\theta^2\beta(\theta+5\beta)}{(\theta+2\beta)(\theta+\beta)}.
\end{align*}
As we see, the formula of $\mbox{dCov}(X,X)$ is complicated for the 2-component exponential mixture distribution. The correlations are given as follows. 
\begin{eqnarray*}
\rho_g(X,Y) 
&=&\frac{p(1-p)(\theta-\beta)^2}{(2p-p^2)\theta^2+(1-p^2)\beta^2+(1-2p+2p^2)\theta\beta},\\
\rho_d(X,Y) &=& \frac{p(1-p)(\theta-\beta)^2}{2(\theta+\beta) \sqrt{\mbox{dCov}(X,X)}},\\
\rho_p^2(X,Y) &=& \frac{p(1-p)(\theta-\beta)^2}{p\theta^2+(1-p)\beta^2+p(1-p)(\theta-\beta)^2}. 
\end{eqnarray*}

Figure \ref{fig:exp} demonstrates Gini correlation, distance correlation and squared Pearson correlation in the exponential mixtures. The cases of $p =0$ or  $p=1$ in (a) and  $\beta=\theta=1$ in (b) have zero Gini, zero distance and zero Pearson correlation coefficients, corresponding to the case of independence of $X$ and $Y$. The value of the Gini correlation is between the squared Pearson correlation and distance correlation. As expected, all correlations increase as the ratio $r = \beta/\theta$ increases for $r \geq 1$.  This result (at least for the Gini and Pearson $R^2$ correlations) can be proved by the positiveness of the  first derivative of the correlation with respect to $r$, which is given in the supplemental file. 

\begin{figure}[h]
\center 
\begin{tabular}{cc}
\includegraphics[width=6cm,height=6cm,keepaspectratio]{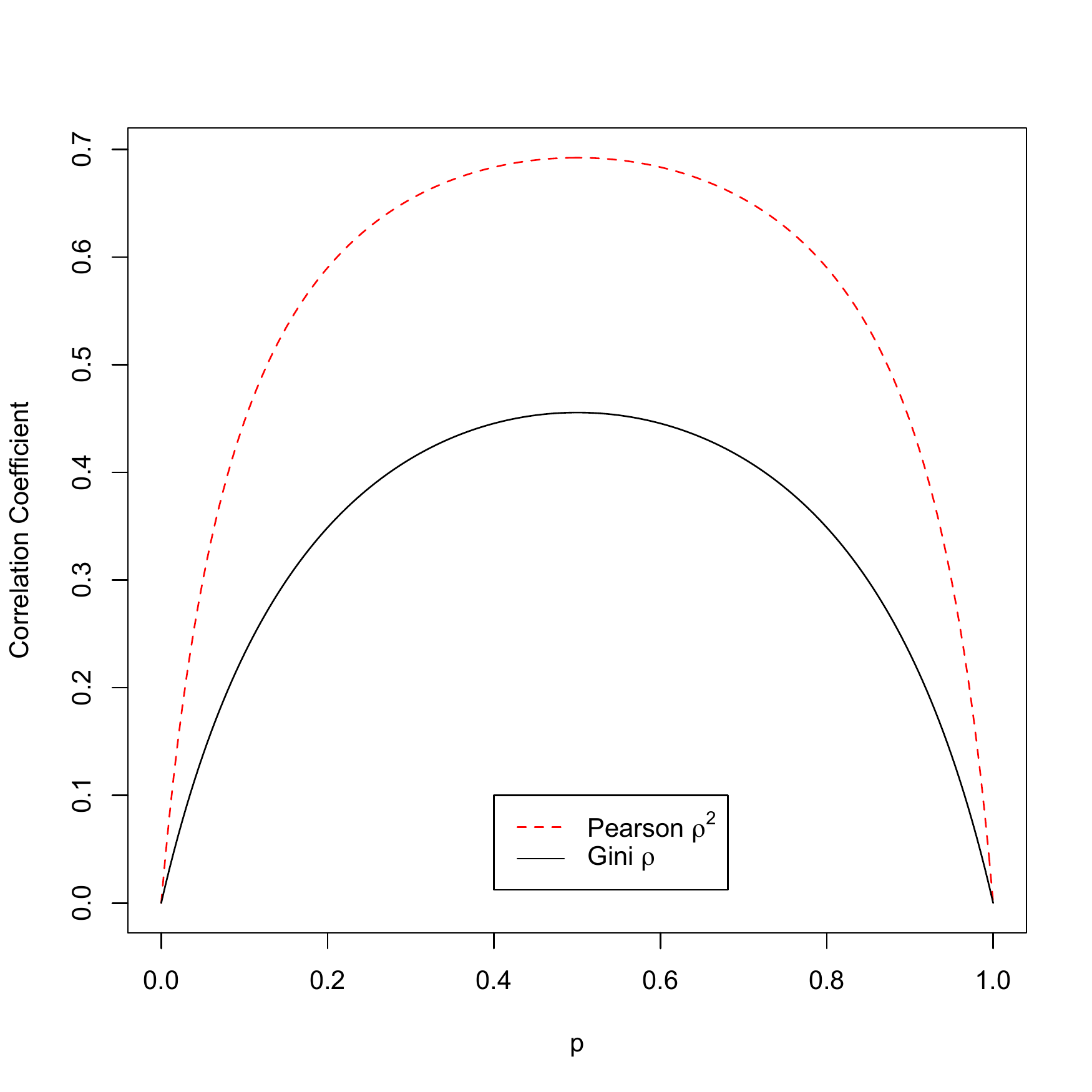}&
\includegraphics[width=6cm,height=6cm,keepaspectratio]{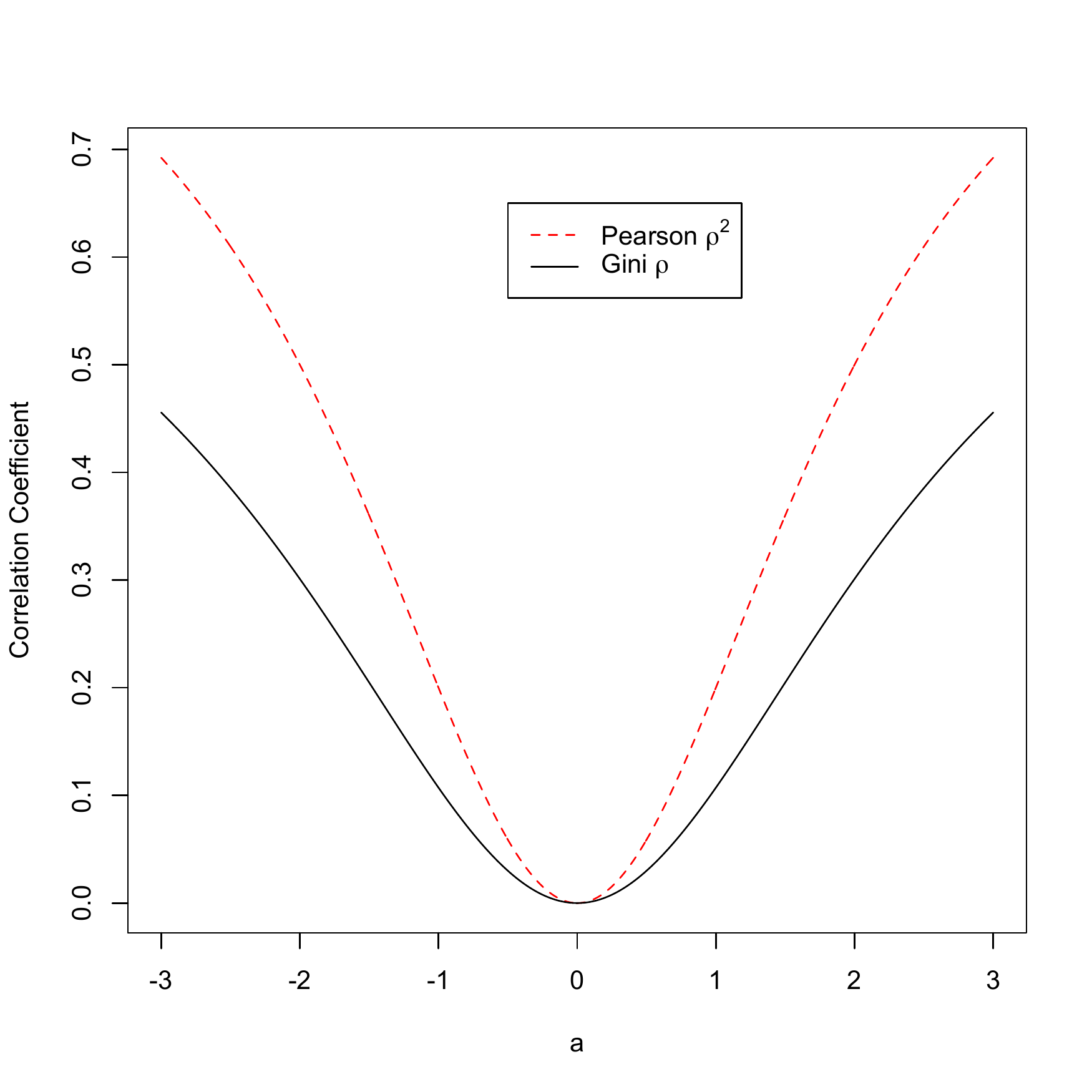} \vspace{-0.05in} \\ 
(a)&(b)
\end{tabular}
\caption{(a) Correlation coefficient vs $p$ in the mixture normal distribution with $a =|\mu_1-\mu_2|/\sigma = 3$; (b) Correlation coefficient vs $a$ with $p=0.5$. \label{fig:norm_mu}}
\end{figure}

\noindent{\bf Example 2.} Let $F_1=\mbox{Normal} (\mu_1, \sigma^2)$, $F_2 = \mbox{Normal}(\mu_2, \sigma^2)$ and $a = |\mu_1-\mu_2|/\sigma$.  We have
$$ \Delta_1 = \Delta_2 = \frac{2\sigma}{\sqrt{\pi}}, \Delta_{12} = \sigma[ 2a \Phi (a/\sqrt{2}) +2\sqrt{2} \phi(a/\sqrt{2})-a], $$
where $\phi(x)$ and $\Phi(x)$ are the density and cumulative functions of the standard normal distribution, respectively.  But it is too complicate to derive formula of $dCov(X,X)$ when $X$ is from a mixture of two normal distributions. In this case, we are only able to derive Gini correlation and the squared Pearson correlation as follows. 
\begin{eqnarray*}
\rho_g(X,Y) &=& \frac{p(1-p)[2a \Phi (a/\sqrt{2}) +2\sqrt{2} \phi(a/\sqrt{2})-a -2/\sqrt{\pi}]}{(p^2+(1-p)^2)/\sqrt{\pi}+p(1-p) [2a \Phi (a/\sqrt{2}) +2\sqrt{2} \phi(a/\sqrt{2})-a]},\\
\rho_p^2(X,Y) &=&  \frac{p(1-p)a^2}{1+p(1-p)a^2}. 
\end{eqnarray*}

For a mixture of two normal distributions with a same standard deviation but different means, independence of $X$ and $Y$ is equivalent to either $p=0, p=1$ in (a) or $a = 0$ in (b) for both correlations, which is demonstrated in Figure \ref{fig:norm_mu}. For  dependence cases, the squared Pearson correlation is larger than the Gini correlation. With any fixed $a \neq 0$ (i.e., $\mu_1 \neq \mu_2$),  the largest correlation is obtained at $p=0.5$ (the balance case) for both correlations.   Also both correlations are monotone increasing functions of $|a|$ for any $p\neq 0$ or $1$.    \\
\begin{figure}[h]
\center 
\begin{tabular}{cc}
\includegraphics[width=6cm,height=6cm,keepaspectratio]{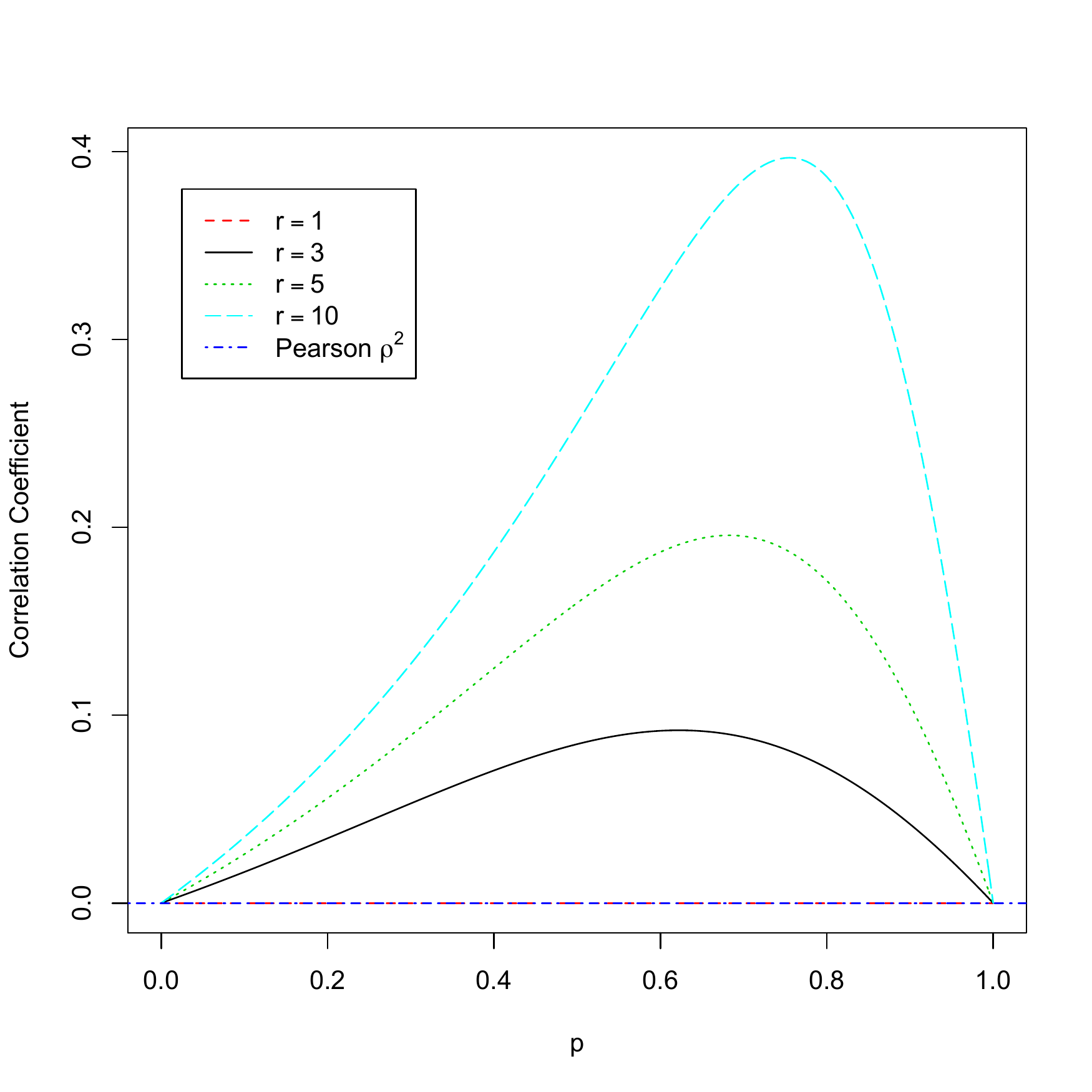}&
\includegraphics[width=6cm,height=6cm,keepaspectratio]{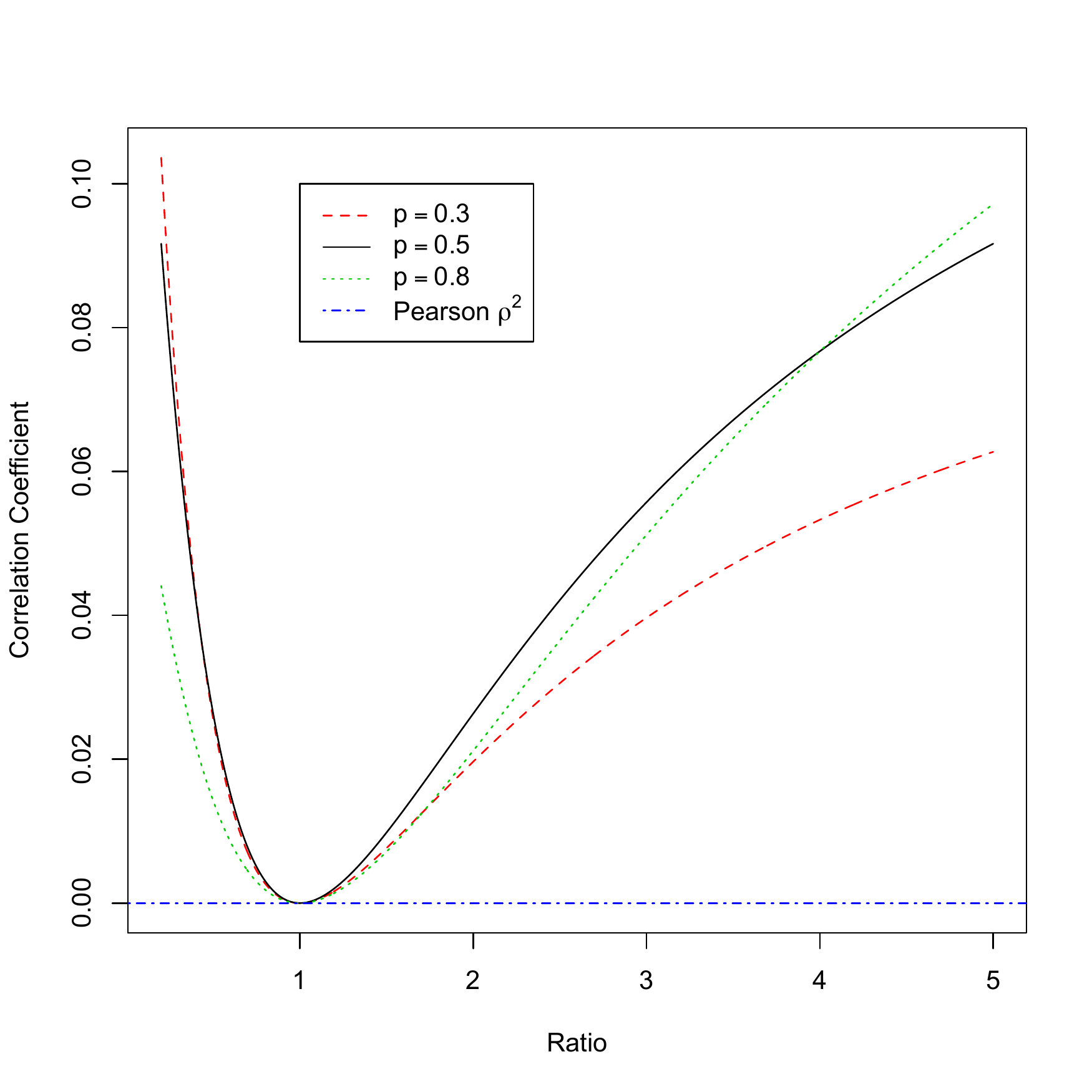} \vspace{-0.05in} \\ 
(a)&(b)
\end{tabular}
\caption{(a) Correlation coefficient vs $p$ in the mixture normal distribution with  $\mu_1=\mu_2$ for different $r =\sigma_2/\sigma_1$; (b) Correlation coefficient vs $r = \sigma_2/\sigma_1$ in the mixture normal distribution with $\mu_1=\mu_2$ for different $p$. \label{fig:norm_sigma}}
\end{figure}

\noindent{\bf Example 3.} Let $F_1 = \mbox{Normal} (\mu, \sigma_1^2)$, $F_2 = \mbox{Normal}(\mu,\sigma_2^2)$ and $r = \sigma_2/\sigma_1$. 
Again, it is too complicate to derive the formula of $\mbox{dCov}(X,X;1)$ in this example.  Since two distributions have a same mean, $\rho_p^2(X,Y)$ is always 0 and hence it completely fails to measure the difference of two distributions when $\sigma_1 \neq \sigma_2$. For the Gini correlation, we have
$$ \Delta_1=\frac{2 \sigma_1}{\sqrt{\pi}}, \Delta_2= \frac{2\sigma_2}{\sqrt{\pi}}, \Delta_{12} = \frac{\sqrt{2(\sigma_1^2+\sigma_2^2)}}{\sqrt{\pi}}. $$
Then 
\begin{eqnarray*}
\rho_g(X,Y) 
&=&\frac{p(1-p)(\sqrt{2(1+r^2)}-1-r)}{p^2+(1-p)^2r+p(1-p)\sqrt{2(1+r^2)}}.
\end{eqnarray*}

Figure \ref{fig:norm_sigma} (a) shows Gini correlation changes with $p$ for normal mixture under different ratios of standard deviations. Figure \ref{fig:norm_sigma} (b) shows the changes of Gini correlation with ratio of standard deviations of normal mixture under different $p$. In the cases of $p=0, 1$ and $r=1$ in (a)  and the case of the ratio to be 1 in (b),  the Gini correlation is 0,  corresponding to the independence of $X$ and $Y$.  The Gini correlation is monotonically increasing in  $r>1$ for each $p \neq 0$  or $1$.

\section{Inference} \label{sec:inference}
\subsection{Estimation}
Suppose a sample data $ {\cal D} =\{(\bi x_i,y_i)\} $  for $i = 1,...,n$ available. We can write $ {\cal D} = \{{\cal D}_1, {\cal D}_2,...,{\cal D}_K \}$, where ${\cal D}_k = \{ \bi x_1^{(k)},...,\bi x_{n_k}^{(k)}\}$ is the sample with $y_i=L_k$ and  $n_k$ is the number of observations in ${\cal D}_k$.  First, $p_k$ is estimated by $n_k/n$. Secondly, $\Delta_k$ and $\Delta$ can be estimated either by U-statistics ($U's$) or V-statistics ($V's$) respectively as follows. 
\begin{align*}
&U's: \;\;\ \hat{\Delta}_k(\alpha)= {n_k \choose 2}^{-1} \sum_{1 \leq i<j \leq n_k} \|\bi x_i^{(k)} -\bi x_j^{(k)}\| ^{\alpha},\;\;\;\;\hat{\Delta}(\alpha)={n \choose 2}^{-1} \sum_{1\leq i<j\leq n} \|\bi x_i -\bi x_j\| ^{\alpha}  \\
&V's: \;\; \hat{\Delta}_k(\alpha) = n_k^{-2}\sum_{1 \leq i,j \leq n_k} \|\bi x_i^{(k)} -\bi x_j^{(k)}\| ^{\alpha},  \;\;\; \hat{\Delta}(\alpha) = n^{-2} \sum_{1\leq i,j \leq n} \|\bi x_i -\bi x_j\| ^{\alpha}
\end{align*}
 Then a corresponding point estimator of $\rho_g(\alpha)$ is 
 \begin{align} \label{rghat}
 \hat{\rho}_g (\alpha)= 1-\frac{\sum_{k=1}^K \hat p_k \hat{\Delta}_k(\alpha)}{\hat{\Delta}(\alpha)}=\frac{\hat{\Delta}(\alpha)- \sum_{k=1}^K \hat p_k \hat{\Delta}_k(\alpha)}{\hat{\Delta}(\alpha)}. 
 \end{align}
 The U statistic and V statistic estimators of $\Delta_k$ and $\Delta$ are only different by the factors of $n_k/(n_k-1)$ and $n/(n-1)$, respectively.  Those factors converge to 1 and hence asymptotic properties of the above $U$ and $V$ estimators holds the same. For the convenience of theoretical limiting distribution developments,  we will focus on the $V$ estimator. 
 We have the following theorems. 

\begin{theorem}\label{consistency}
If $\E \|\bi X\|^\alpha <\infty$ and $p_k >0$ for all $k=1,...,K$, then almost surely 
$$ \lim_{n\rightarrow \infty} \hat{\rho}_g(\alpha)  = \rho_g(\alpha).$$
\end{theorem}
{\bf Proof.} By the SLLN,  $\hat{p}_k = \frac{1}{n} \sum_{i=1}^n I(y_i = L_k)$ converges to $p_k$ with probability 1. 
Also by the almost sure behavior of $V$-statistics (p. 228, Serfling (1980)), $\hat{\Delta}_k(\alpha)$ and  $\hat{\Delta}(\alpha)$ converge with probability 1 to $\Delta_k(\alpha)$ and $\Delta(\alpha)$, respectively. Let $g$ be the function $g(a_1,...,a_K, b_1,...,b_K, b) = 1-\sum_{k=1}^K a_kb_k /b$,  which is continuous for $b>0$. Therefore, the strong consistency of the sample Gini correlation follows by the fact that $\hat{\rho}_g(\alpha) = g(\hat{p}_1,..., \hat{p}_K, \hat{\Delta}_1(\alpha),...,\hat{\Delta}_K(\alpha),  \hat{\Delta}(\alpha))$.  \hfill{$\square$}
 
\begin{theorem}\label{thmn0}
Suppose that $\E \|\bi X\|^{2\alpha}<\infty$,  $p_k >0$ for all $k=1,...,K$ and $\rho_g(\alpha)\neq 0$.   We have
$$ \sqrt{n} (\hat{\rho}_g(\alpha) -\rho_g(\alpha)) \rightarrow N(0, \sigma_g^2(\alpha)),$$
where $\sigma_g^2(\alpha)$ is the asymptotic variance given in the proof. 
\end{theorem}

\noindent{\bf Proof}. For simplicity of presentation, we suppress $\alpha$ in notations in the proof without confusion.  Let $\bi q$ be the vector of length $K(K-1)$ with elements $\{p_kp_l\}_{1\leq k \neq l \leq K}$. 
Let  $\bi h$ be the kernel functions of form $\bi h = \{h_{kl}\}_{1\leq k \neq l \leq K}$, where  
$$h_{kl}:=h(\bi x_1^{(k)}, \bi x_2^{(k)}; \bi  x_1^{(l)},\bi x_2^{(l)})=\frac{1}{2}[\|\bi x_1^{(k)}-\bi x_1^{(l)} \|^{\alpha}+\|\bi x_2^{(k)}-\bi x_2^{(l)} \|^{\alpha} - \|\bi x_1^{(k)}-\bi x_2^{(k)} \|^{\alpha}-\|\bi x_1^{(l)}-\bi x_2^{(l)}\|^{\alpha}].$$ 
Let $(\bi X_1^{(k)}, \bi X_2^{(k)})$ and $(\bi X_1^{(l)}, \bi X_2^{(l)})$ be independent pairs independently from distributions $F_k$ and $F_l$, respectively.  Then $\bi q^T \E \bi h(\bi X_1^{(1)}, \bi X_2^{(1)};  ...;  \bi X_1^{(K)}, \bi X_2^{(K)})= \sum_{1\leq k \neq l \leq K} p_kp_lT(\bi X^{(k)},\bi X^{(l)};\alpha)$, which is gCov$(\bi X,Y;\alpha)$ by Remark \ref{rem:2-4}. Let  $\hat{T}_{kl}$ be the V-statistic estimator of $T(\bi X^{(k)},\bi X^{(l)};\alpha)$. That is,  
\begin{eqnarray*}
\hat{T}_{kl} &=& \frac{1}{n_k^2}\frac{1}{n_l^2}\sum_{i_1,i_2=1}^{n_k}\sum_{j_1,j_2=1}^{n_k} h(\bi x_{i_1}^{(k)}, \bi x_{i_2}^{(k)}, \bi x_{j_1}^{(l)}, \bi x_{j_2}^{(l)} )\\
&=&\frac{1}{n_kn_l} \sum_{i_1=1}^{n_k}\sum_{j_1=1}^{n_l}   \| \bi x^{(k)}_{i_1} - \bi x^{(l)}_{j_1}\|^\alpha- \frac{1}{2n_k^2} \sum_{i_1,i_2=1}^{n_k} \| \bi x^{(k)}_{i_1} - \bi x^{(k)}_{i_2}\|^\alpha -\frac{1}{2n_l^2} \sum_{j_1,j_2=1}^{n_l} \| \bi x^{(l)}_{j_1} - \bi x^{(l)}_{j_2}\|^\alpha. 
\end{eqnarray*}
Then  an estimator of $gCov(\bi X,Y;\alpha)$ given by $\sum_{1\leq k \neq l \leq K} \hat{p}_k\hat{p}_l \hat{T}_{kl}$ is same as the $V$-statistic estimator given in (\ref{rghat}) because
\begin{align*}
&\sum_{1\leq k \neq l \leq K} \hat{p}_k\hat{p}_l \hat{T}_{kl} = \frac{1}{n^2} \left[\sum_{1\leq k \neq \leq K}\sum_{i_1=1}^{n_k}\sum_{j_1=1}^{n_l}   \| \bi x^{(k)}_{i_1} - \bi x^{(l)}_{j_1}\|^\alpha - \sum_{k=1}^K \frac{n-n_k}{n_k} \sum_{i_1,i_2=1}^{n_k} \| \bi x^{(k)}_{i_1} - \bi x^{(k)}_{i_2}\|^\alpha\right]\\
&=\frac{1}{n^2}\left[ \sum_{k=1}^K \sum_{i_1,i_2=1}^{n_k}\| \bi x^{(k)}_{i_1} - \bi x^{(k)}_{i_2}\|^\alpha +  \sum_{k \neq l}\sum_{i_1=1}^{n_k}  \sum_{j_1=1}^{n_l}  \| \bi x^{(k)}_{i_1} - \bi x^{(l)}_{j_1}\|^\alpha    \right]- \sum_{k=1}^K \frac{n_k}{n} \frac{1}{n_k^2}\sum_{i_1,i_2=1}^{n_k}\| \bi x^{(k)}_{i_1} - \bi x^{(k)}_{i_2}\|^\alpha  \\
&=\hat{\Delta}(\alpha)-\sum_k \hat{p}_k \hat {\Delta}_k(\alpha).
\end{align*}
Consider the centered kernel function $\tilde{h}_{kl} ={h}_{kl}-\E h_{kl}$ and its first order projections as follows. 
\begin{align}
&\tilde{h}_{kl}^{10} (\bi x^{(k)}) = \E \tilde{h}_{kl}(\bi x_1^{(k)}, \bi X_2^{(k)}; \bi X_1^{(l)},\bi X_2^{(l)}) =\frac{1}{2}[\E \|\bi x^{(k)}-\bi X_1^{(l)}\|^\alpha - \E  \|\bi x^{(k)}-\bi X_2^{(k)}\|^\alpha],\nonumber\\
&\tilde{h}_{kl}^{01} (\bi x^{(l)}) =  \E \tilde{h}_{kl}(\bi X_1^{(k)}, \bi X_2^{(k)}; \bi x_1^{(l)},\bi X_2^{(l)})= \frac{1}{2}[\E \|\bi X^{(k)}-\bi x^{(l)}\|^\alpha - \E  \|\bi x^{(l)}-\bi X_2^{(l)}\|^\alpha].  \label{1proj}
\end{align}
Denote  the first order projection of $\hat{T}_{kl}$ as $\hat{T}_{kl}^{(1)} = 2 [\frac{1}{n_k} \sum_{i=1}^{n_k} \tilde{h}_{kl}^{10} (\bi x_i^{(k)})+\frac{1}{n_l} \sum_{j=1}^{n_l} \tilde{h}_{kl}^{01} (\bi x_j^{(l)})]$. Then 
\begin{align*}
\sum_{k \neq l} \hat{p}_k\hat{p}_l \hat{T}_{kl}^{(1)} = \frac{4}{n^2}\sum_{k \neq l} n_l \sum_{i=1}^{n_k}\tilde{h}_{kl}^{10} (\bi x_i^{(k)}).
\end{align*}
If at least one $\sigma^2_{kl} (\alpha):=var(\tilde{h}_{kl}^{10} (\bi X^{(k)}))$ is  not zero,  the  corresponding V-statistic  has an asymptotic normal distribution (Theorem A of Section 6.4, Serfling(1980)).  That is, 
$$\sqrt{n}( \hat{\Delta}(\alpha)-\sum_k \hat{p}_k \hat {\Delta}_k(\alpha) -\mbox{gCov}(\bi X,y;\alpha))  \rightarrow {\cal N}(0, 16 \sum_{k\neq l}^K p_l^2p_k \sigma_{kl}^2)(\alpha). $$ 
Next, we need to show that $\rho_g(\bi X, Y;\alpha) = 0$ if and only if all $\sigma^2_{kl}(\alpha) =0$. First,  $\rho_g(\bi X, Y;\alpha) = 0$ implies that $F_1=F_2...=F_K=F$,  $\tilde{h}_{kl}^{10} (\bi X^{(k)}))=0$  and hence all $\sigma^2_{kl}(\alpha) =0$. On the other hand, if $\sigma^2_{kl}(\alpha) =0$, then $\tilde{h}_{kl}^{10} (\bi X^{(k)}))$ is a constant $C$ almost surely. Taking  the expectation on $\tilde{h}_{kl}^{10} (\bi X^{(k)}))$ gives $C=0$. Also $\E \tilde{h}_{kl}^{10} (\bi X^{(k)})) = \Delta_k(\alpha) -\Delta_{kl}(\alpha) $, which is equal to $\E \tilde{h}_{lk}^{10} (\bi X^{(l)}))=\Delta_l(\alpha)-\Delta_{kl}(\alpha)$.  Then we obtain $\Delta_1(\alpha) =\Delta_2(\alpha)=...=\Delta_K(\alpha) =\Delta(\alpha)$.  That implies $\rho_g(\bi X, Y;\alpha) = 0$. 
Hence,  if $\rho_g(\bi X, Y;\alpha) > 0$, there is at least one nonzero $\sigma^2_{kl}$.  
Then by Slustky's theorem, $$\sqrt{n}(\hat{\rho}_g(\alpha) - \rho_g(\bi X,Y; \alpha)) \rightarrow {\cal N}(0, \sigma_g^2(\alpha)), $$
where $\sigma_g^2(\alpha) = 16 \sum_{k\neq l}^K p_l^2p_k \sigma_{kl}^2(\alpha)/\Delta^2(\alpha)$. 
\hfill{$\square$}

Although we have a formula of $\sigma_g^2(\alpha)$, it is difficult to calculate in practice because it depends on unknown $F_k$ and $p_k$. To overcome this difficulty, we can estimate $\sigma^2_g(\alpha)$ by the jackknife method.
Let $\hat {\rho}_{(-i)}(\alpha)$ be the Gini correlation estimator based on the sample with the $i^{th}$ observation deleted. Then the jackknife estimator of standard error $\sigma_g(\alpha)/\sqrt{n}$ is
\begin{align}\label{jelrho}
\hat{SE}(\alpha)=\sqrt{\frac{n-1}{n}\sum_{i=1}^n ({\hat{\rho}}_{(-i)}(\alpha) -\bar{\hat{\rho}}_{(\cdot)}(\alpha))^2},
\end{align}
where $\bar{\hat{\rho}}_{(\cdot)}(\alpha)=1/n \sum_{i=1}^n {\hat{\rho}}_{(-i)}(\alpha)$. See Shao and Tu (1996) for details. Then a $(1-\gamma) 100\%$ confidence  interval of $\rho_g(\alpha)$ is 
$$ \hat{\rho}_g(\alpha) \pm z_{\gamma/2} \hat{SE}(\alpha),$$
where $z_{\gamma/2}$ is the $(1-\gamma/2)100\%$ quantile of the standard normal variable. 

Theorem \ref{thmn0} states the asymptotic normality of $\hat{\rho}_g(\alpha)$ when $\bi X$ and $Y$ are dependent. However, if $\rho_g(\alpha) =0$ when $\bi X$ and $Y$ are independent,  the behavior of  $\hat{\rho}_g(\alpha)$ is quite different since $\sigma_g^2=0$.  In this degenerate case, the limiting distribution of $n(\hat{\Delta}(\alpha) -\sum_k \hat{p}_k \hat{\Delta}_k(\alpha)) $ converges in distribution to a quadratic form of Gaussian random variables,  same as the result in Sz\'{e}kely and Rizzo (2005, 2013) and Rizzo and Sz\'{e}kely (2010). They have proved for balanced cases that $S_{\alpha}$, the between sample dispersion by the DISCO decomposition, converges in distribution to a quadratic form of centered Gaussian random variables.   We state the following theorem and provide a proof that does not require balance sizes. 

\begin{theorem} \label{thm0}
If $\rho_g(\bi X,Y; \alpha)=0$ and $\E \|\bi X\|^{2\alpha}<\infty$, then 
$$ n \hat{\rho}_g(\alpha)   \rightarrow \frac{4}{\Delta(\alpha)}\left [ \sum_{s=1}^\infty \sum_{k=1}^K (1-p_k)  \lambda_s Z_{s,k}^2+ \sum_{s=1}^\infty \sum_{1\leq k <l\leq K} \sqrt{p_kp_l} \lambda_s Z_{s,k}Z_{s,l}\right ] ,$$
where $Z_{s,k} (k=1,...,K, s=1,2,...)$ are independent standard normal variates and $\lambda_s$ are nonnegative coefficients. 
\end{theorem} 

\noindent {\bf Proof}. Under $\rho_g(\bi X,Y; \alpha)=0$, $F_1=F_2=...=F_K=F$ implies all $\tilde{h}_{kl}^{10} (\bi x^{(k)})$'s in (\ref{1proj}) are zero almost surely. In this degenerate case, we need the second order projections of $\tilde{h}_{kl}$. 
\begin{eqnarray*}
\tilde{h}_{kl}^{20} (\bi x_1^{(k)}, \bi x_2^{(k)}) &=& \E \tilde{h}_{kl}(\bi x_1^{(k)}, \bi x_2^{(k)}; \bi X_1^{(l)},\bi X_2^{(l)}) \\
&=&\frac{1}{2}[\E \|\bi x_1^{(k)}-\bi X_1^{(l)}\|^\alpha + \E \|\bi x_2^{(k)}-\bi X_2^{(l)}\|^\alpha -\|\bi x_1^{(k)}-\bi x_2^{(k)}\|^\alpha-\Delta_l(\alpha)]  \\
\tilde{h}_{kl}^{02} (\bi x_1^{(l)}, \bi x_2^{(l)}) &=& \E \tilde{h}_{kl}(\bi X_1^{(k)}, \bi X_2^{(k)}; \bi x_1^{(l)},\bi x_2^{(l)}) \\
&=&\frac{1}{2}[\E \|\bi X_1^{(k)}-\bi x_1^{(l)}\|^\alpha + \E \|\bi X_2^{(k)}-\bi x_2^{(l)}\|^\alpha -\|\bi x_1^{(l)}-\bi x_2^{(l)}\|^\alpha-\Delta_k(\alpha)], \\
\tilde{h}_{kl}^{11} (\bi x_1^{(k)}, \bi x_1^{(l)}) &=& \E \tilde{h}_{kl}(\bi x_1^{(k)}, \bi X_2^{(k)}; \bi x_1^{(l)},\bi X_2^{(l)})\\
&=&\frac{1}{2}[ \|\bi x_1^{(k)}-\bi x_1^{(l)}\|^\alpha +\Delta_{kl}(\alpha) - \E \|\bi x_1^{(k)}-\bi X_2^{(k)}\|^\alpha - \E  \|\bi x_1^{(l)}-\bi X_2^{(l)}\|^\alpha].
\end{eqnarray*}
If $\rho_g(\bi X,Y; \alpha)=0$,  we have $\tilde{h}_{kl}^{20}  = \tilde{h}_{kl}^{02} = \tilde{h}_{kl}^{11} $ for all $k \neq l$ and denote them as $h_2$. 
Let $h_2(\bi x_1, \bi x_2) = \sum_{s=1}^\infty\lambda_s \phi_s(\bi x_1) \phi_s(\bi x_2)$, where 
$$ \int_{\mathbb{R}^d} h_2(\bi x_1, x_2) \phi_s(\bi x_2) dF(\bi x_2) = \lambda_s \phi_s(\bi x_1). $$
Under the assumption of $\E \|\bi X\|^{2\alpha}<\infty$,  we have $\sum_{s=1}^\infty \lambda_s < \infty$. 

Denote the second order projection of $\hat{T}_{kl}$ as $\hat{T}_{kl}^{(2)}$ that is given by
$$\hat{T}_{kl}^{(2)} = \frac{2}{n_k^2} \sum_{1\leq  i,j \leq n_k} h_2(\bi x_i^{(k)}, \bi x_j^{(k)})+ \frac{2}{n_l^2}\sum_{1\leq  i,j\leq n_l} h _2 (\bi x_i^{(l)}, \bi x_j^{(l)}) + \frac{4}{n_kn_l}\sum_{i=1}^{n_k}\sum_{j=1}^{n_l} h_2(\bi x_i^{(k)}, \bi x_j^{(l)}). $$
We oobtain the second order projection of the estimator to be
\begin{align*}
&\sum_{k \neq l} \hat{p}_k\hat{p}_l \hat{T}_{kl}^{(2)} =\frac{1}{n} [\sum_{k=1}^K 4(1-p_k) \frac{1}{n_k} \sum_{1\leq i,j \leq n_k} h_2(\bi x_i^{(k)}, \bi x_j ^{(k)}) 
+ \sum_{k \neq l} 4 \sum_{i=1}^{n_k}\sum_{j=1}^{n_l} h_2(\bi x_i^{(k)}, \bi x_j^{(l)})]. 
\end{align*}
By the $V$-statistic theorem (Theorem B of Section 6.4, Serfling (1980)), we have 
$$ n( \hat{\Delta} - \sum_k \hat p_k \Delta_k) \rightarrow \sum_{s=1}^\infty \sum_{k=1}^K 4(1-p_k)  \lambda_s Z_{s,k}^2+ \sum_{s=1}^\infty \sum_{1\leq k <l\leq K} 4\sqrt{p_kp_l} \lambda_s Z_{s,k}Z_{s,l} $$
where $Z_{s,k} (k=1,...,K, s=1,2,...)$ are independent standard normal variates. An immediate application of Slustky's Theorem completes the proof. \hfill{$\square$}

 \subsection{Testing Dependence}     
The Gini correlation is zero if and only $\bi X$ and $Y$ are independent. Hence, for a given $0 < \alpha<2$, the independence test can be stated as

\begin{equation}\label{test}
 {\cal H}_0: \rho_g(\bi X, Y; \alpha) = 0,\;\;\;\; \mbox{vs}\;\;\;\; {\cal H}_1: \rho_g(\bi X, Y; \alpha)=\rho_0 >  0. 
 \end{equation}
Reject ${\cal H}_0$ if $\hat{\rho}_g$ is large. The critical value of the test of significance level $\gamma$, however, is difficult to obtain from Theorem \ref{thm0} by two reasons. Firstly $\lambda_l$'s depend on distribution $F$, which is usually unknown.  Secondly, it is a mixture of infinitely many distributions. To overcome this difficulty, a permutation procedure is used to estimate the critical value and p-value.  Let $\nu = 1: n$ be the vector of original sample indices of the sample for $Y$ labels and $\hat{\rho}_g(\alpha) = \hat{\rho}(\nu;\alpha)$. Let $\pi(\nu)$ denote a permutation of the elements of $\nu$ and the corresponding $\hat{\rho}_g(\pi;\alpha)$ is computed. Under the ${\cal H}_0$, $\hat{\rho}_g(\nu)$
and $\hat{\rho}_g(\pi;\alpha)$ are identically distributed for every permutation $\pi$ of $\nu$. Hence, based on $M$ permutations,  the critical value $q_{\gamma}$ is estimated by the $(1-\gamma)100\%$ sample quantile of $\hat{\rho}_g(\pi_m;\alpha)$, $m=1,...,M$ and the p-value is estimated by the proportion of $\hat{\rho}_g(\pi_m;\alpha)$ greater than $\hat{\rho}_g(\nu;\alpha)$. Usually $100\leq M\leq 1000$ is sufficient for a good estimation on the critical value or p-value.  In the simulation next section, we use $M=200$. Further, if $\rho_0$ is specified, the power of the test can be computed by
$1- \Phi((q_{\gamma}-\rho_0)/(\hat{v}_g/\sqrt{n } ))$, where $\Phi(x)$ is the cumulative distribution function of the standard normal random variable. 


\subsection{Computation issues}
The computation of the sample Gini correlation $\hat{\rho}_g(\alpha)$ in \eqref{srhog} is straightforward. In general, it has a computational complexity $O(n^2)$ since all distinct pair distances need to calculate. However, when the numerical variable is univariate and $\alpha =1$, we have a much faster algorithm that only costs $O(n \log n)$ computation.   This is because the univariate Gini mean distance can be written as a linear combination of order statistics \cite{Schechtman87}. Suppose that $x_{(1)} \leq x_{(2)} \leq \cdots \leq x_{(n)}$ are the order statistics of $x_1, x_2, ..., x_n$. Then 
$$ \hat{\Delta}(X; 1) = {n \choose 2}^{-1} \sum_{1\leq i<j\leq n} | x_i - x_j| = {n \choose 2}^{-1} \sum_{i=1}^n (2i-n-1) x_{(i)}. $$
This fast algorithm is crucial for Gini correlation in application of feature screening. For the classification problem with ultrahigh-dimensional data, the first step is to screen out unimportant predictors. We can rank features by their Gini correlations with the class label and screen out less correlated predictors, analogue to the sure independent screening procedures  (Fan and Lv 2008, Li, Zhong and Zhu 2012)  in which they consider the response variable is numerical and use Pearson correlation or distance correlation to do feature selection. 

For sample distance correlation $\hat \rho_d(\alpha)$, its computation follows as the average of the element-wise product of two centered pairwise distance matrices, which is described in \cite{Szekely07}. With small adjustments in centering, an unbiased estimator is provided in \cite{Szekely04}. 
More specifically, let $A=(a_{ij})$ be a symmetric, $n \times n$, centered distance matrix of sample $\bi x_1,\cdots, \bi x_n$. The $(i,j)$-th entry of $A$ is
\begin{equation*}
    A_{ij} = 
    \begin{cases}
    a_{ij}-\frac{1}{n-2}a_{i\cdot}-\frac{1}{n-2}a_{\cdot j} + \frac{1}{(n-1)(n-2)}a_{\cdot \cdot}, & i\ne j;\\
    0, & i=j,
    \end{cases}
\end{equation*}
where $a_{ij} = \|\bi x_i-\bi x_j\|^{\alpha}$, $a_{i\cdot} = \sum_{j=1}^n a_{ij}$, $a_{\cdot j} = \sum_{i=1}^n a_{ij}$, and $a_{\cdot \cdot}=\sum_{i,j=1}^n a_{ij}$. Similarly, using the set difference metric, a symmetric, $n \times n$, centered distance matrix is calculated for samples $y_1,\cdots, y_n$ and denoted by $B = (b_{ij})$. Unbiased estimators of $dCov(\bi X,Y;\alpha)$ and $dCov(\bi X, \bi X;\alpha)$ are given  respectively   as,  
\begin{align*}
   &\frac{1}{n(n-3)}\sum_{i\ne j}A_{ij}B_{ij},\;\;\; \;\;\frac{1}{n(n-3)}\sum_{i\ne j}A_{ij}^2. 
\end{align*}

Note that for univariate $X$,  a fast $O(n \log n)$ algorithm for sample distance correlation is available \cite{Huo16}, but the implementation is tricky due to dealing with the centering process. 

Another computation issue is the choice of $\alpha$, the parameter of distance metric in $\mathbb{R}^d$. A natural choice is $ \alpha=1$, which corresponds to the Euclidean distance and leads to fast algorithms for the univariate case. However, if outliers appear in data, we may choose a small $\alpha$ value so that the Gini and distance correlations are insensitive to the outliers,  as mentioned in Remark \ref{alpha}. We can also choose the $\alpha$ value to maximize the correlations. The idea  is similar to the approach in Sarmanov (1958) and Zhang et al. (2019). They choose the transformation of the data to achieve the largest association.  We can select the metric on the original data so that the correlation is the greatest. It is worthwhile to continue the research in this directions in the future.  In the next section, we use $\alpha=1$ in the first three simulation studies and in the real data application. And $\alpha=0.5, 0.75, 1$ are used in the last simulation for demonstration of different $\alpha$ values for a heavy-tailed distribution. 

\section{Experiment}
\label{sec:experiment}
\subsection{Simulation}
Four simulations are conducted to demonstrate the performance of Gini correlation. The first one is to check the coverage probabilities of the confidence intervals based on the asymptotic normality with the asymptotic variance estimated by the Jackknife method.   The second simulation is to compare dependence tests based on Gini correlation and the distance correlation. The third one compares computation time of Gini correlation and the distance correlation and the last simulation is to illustrate that a small $\alpha$ value is more proper for data from heavy-tailed distributions. 

\begin{table}[b]
\centering
\caption{Coverage probabilities of the confidence intervals. \label{tab:ci}}
\begin{tabular}{l|c|cc|cr} \hline\hline
\multirow{2}{*}{Distribution}&\multirow{2}{*}{Parameter}&\multicolumn{2}{c|}{Level $=0.90$} & \multicolumn{2}{c}{Level $=0.95$} \\  \cline{3-6}
&& $n=60$ & $n=120$&$n=60$ & $n=120$\\ \cline{1-6}
\multirow{2}{*}{$0.5\mbox{Exp}(1)+0.5\mbox{Exp}(4)$}& $\rho_g=0.1525$ &0.9031&0.9007&0.9442&0.9472 \\
& $\rho_d=0.1191$& 0.5059&0.7690&0.6771&0.8802\\ \hline
$0.5 {\cal N}(0,1)+0.5 {\cal N}(3,1)$& $\rho_g=0.4556$ & 0.8898&0.8934&0.9323&0.9437\\  \hline
 $0.5 {\cal N}(0,1)+0.5 {\cal N}(0,3^2)$& $\rho_g=0.0557$& 0.9201&0.9093& 0.9531& 0.9524\\
\hline \hline 
\end{tabular}
 \end{table}

For the first simulation on confidence intervals, we consider examples studied in the previous section. The coverage probabilities of confidence intervals in Table \ref{tab:ci} are computed based on $10000$ repetitions.  Two confidence levels 0.90 and 0.95 are considered under sample sizes of $n=60$ and $n=120$. Comparison with confidence intervals of $\rho_d$ is only available for Example 1 where random samples are generated from the mixture of two exponential distributions. This is because the true values of $\rho_d$ are unknown in the other two cases.

From Table \ref{tab:ci}, we observe that the coverage probabilities of confidence intervals of $\rho_d$ are unacceptable. One possible reason is the double-centering procedure in the computation of the sample distance correlation, which makes its finite sample performance undesirable.  The coverage probabilities of confidence intervals for $\rho_g$ are satisfying. They are reasonably close to the nominal levels even under sample size of $n=60$.  

For the second simulation on dependence test, three methods are compared.
The following three scenarios with unbalanced $\bi p = (p_1,p_2,p_3)=(1/4,1/3, 5/12)$ and balanced $\bi p=(1/3,1/3,1/3)$ of the total sample sizes of $(n=60, n=120)$ are considered. 
\begin{itemize}
\item $X\sim p_1 \exp(1)+ p_2 \exp(\theta_1)+p_3\exp(\theta_2)$;
\item $X\sim p_1 {\cal N}(0,1)+ p_2 {\cal N}(\mu_1,1)+p_3{\cal N}(\mu_2,1)$;
\item $X\sim p_1 {\cal N}(0,1)+ p_2 {\cal N}(0,\sigma_1)+p_3{\cal N}(0,\sigma_2)$.
\end{itemize}
The size and power of each test based on 1000 repetitions are reported in Table \ref{tab:power}. The cases of $\theta_1=\theta_2=1$, $ \mu_1=\mu_2=0 $ and $\sigma_1=\sigma_2=1$ imply independence of $X$ and $Y$. Two  tests maintain the test level 0.05 well. For the unbalanced cases, the $\rho_g$ test is slightly more powerful than the $\rho_d$ test. For normal mixtures with different locations, performance of two tests are similar with the $\rho_g$ test slightly better.  The power of the $\rho_g$ test is about 2\%-3\% higher than the $\rho_d$ test in exponential mixtures and in normal mixtures with different scales. Power of the tests in balanced cases is higher than that in unbalanced cases.  In the balanced case, two permutation tests  are asymptotically equivalent since the population Gini correlation is a multiple of the population distance correlation (Remark \ref{balance}). This is demonstrated in the simulation with $\bi p = (1/3,1/3,1/3)$. The powers of  the $\rho_g$ method are very similar to the powers of the $\rho_d$ test in balanced scenarios.  From Remark \ref{rem:2class}, we also have the two permutation tests on the $K=2$ problem asymptotically equivalent. However, the Gini correlation method is preferred as it has lower empirical computation time even though the two correlations have the same computation complexity.
\begin{table}[H]
\caption{Size and power of dependence tests at 0.05 significance level.  \label{tab:power}} 
\centering
\scalebox{0.90}{
\begin{tabular}{lccc|cccccr} \hline\hline
Dist&$\bi p$ &$n$ & Method & \multicolumn{6}{c}{$(\theta_1,\theta_2)$} \\ 
&&&&(1,1) &(1.2,1.4)&(1.4,1.8)&(1.6,2.2)&(1.8,2.6)&(2,3)\\ \hline
\multirow{8}{*}{Exp}& $(\frac{1}{4}, \frac{1}{3}, \frac{5}{12})$& 60 & $\rho_g$& .057 &.135&.280&.470&.655&.758\\
&&&$\rho_d$ & .058& .132&.264&.435&.614&.728\vspace{0.1cm}\\ 

&& 120& $\rho_g$ & .057 &.201&.545&.799&.944&.982\\
&&&$\rho_d$& .055&.195&.522&.777&.924&.975\vspace{0.1cm}\\

&$(\frac{1}{3}, \frac{1}{3},\frac{1}{3})$&60 & $\rho_g$ &.049&.131&.300&.486&.687&.813\\
&&&$\rho_d$ & .049&.131&.302&.481&.688&.814\vspace{0.1cm} \\
& & 120 & $\rho_g$ &.052 &.211&.582&.845&.951&.988\\
&&&$\rho_d$ & .053& .208&.581&.844&.951&.990\\  \hline

&&&&\multicolumn{6}{c}{$(\mu_1,\mu_2)$}\\
&&&&(0,0) &(0.2,0.4)&(0.4,0.8)&(0.6,1.2)&(0.8,1.6)&(1,2)\\ \hline
\multirow{8}{*}{Norm}& ($\frac{1}{4}, \frac{1}{3}, \frac{5}{12}$)& 60 & $\rho_g$ & .058 &.170&.545&.900&.989&1.000\\
&&&$\rho_d$  & .061& .165&.536&.892&.987&1.000\vspace*{0.1cm}\\ 
&& 120& $\rho_g$ & .050 &.306&.857&1.000&1.000&1.000\\
&&&$\rho_d$ & .052&.295&.854&.992&1.000&1.000\vspace{0.1cm}\\
&$(\frac{1}{3}, \frac{1}{3},\frac{1}{3})$&60 & $\rho_g$ &.058&.178&.569&.909&.995&1.000\\
&&&$\rho_d$ & .061&.180&.570&.914&.996&1.000\vspace{0.1cm}\\
& & 120 & $\rho_g$ &.060 &.332&.880&.996&1.000&1.000\\
&&&$\rho_d$ & .059& .331&.875&.994&1.000&1.000\vspace{0.1cm}\\ \hline
&&&&\multicolumn{6}{c}{$(\sigma_1,\sigma_2)$}\\
&&&&(1,1) &(1.2,1.4)&(1.4,1.8)&(1.6,2.2)&(1.8,2.6)&(2,3)\\ \hline
\multirow{8}{*}{Norm} & ($\frac{1}{4}, \frac{1}{3}, \frac{5}{12}$)& 60 & $\rho_g$ & .052 &.066&.114&.202&.310&.438\\
&&&$\rho_d$ & .056& .072&.117&.200&.301&.410\vspace{0.1cm}\\ 
&& 120& $\rho_g$ & .051 &.103&.272&.546&.770&.924\\
&&&$\rho_d$ & .051&.103&.262&.518&.740&.892\vspace{0.1cm}\\
&$(\frac{1}{3}, \frac{1}{3},\frac{1}{3})$&60 & $\rho_g$ &.048&.084&.155&.237&.401&.535\\
&&&$\rho_d$ & .047&.085&.150&.231&.390&.536\vspace{0.1cm}\\
& & 120 & $\rho_g$ &.056 &.111&.302&.640&.844&.956\\
&&&$\rho_d$ & .056& .109&.293&.639&.839&.956\\ \hline\hline
\end{tabular}}
\end{table}

In the next, we conduct a simulation to compare computation time of the Gini correlation and distance correlation. We simulate standard normal random samples in $\mathbb{R}^d$ ($d =1, 10, 100$) of sizes $n =100, 1000, 10000$. Half of sample points are randomly assigned to Class 1 and the other half forms Class 2. The process is repeated 100 times. We use ``dcovU\_stat" function in ``energy" package to compute sample distance correlation. The code is run in a MacPro laptop with one 4-core 2.8 GHz processor.    The mean and standard deviation of the computation time for the Gini correlation and distance correlation are recorded in Table \ref{tab:computation}.

\begin{table}[H]
\center
\begin{tabular}{l|cc|cc|cc} \hline\hline
& \multicolumn{2}{c}{$d=1$} & \multicolumn{2}{c}{$d=10$} &\multicolumn{2}{c}{$d=100$}\\ \cline{2-7}
$n$ & $\rho_g$&$\rho_d$&$\rho_g$&$\rho_d$&$\rho_g$&$\rho_d$\\ \hline
100 & .000(.000)&.001(.002)&.000(.000)&.001(.001)&.001(.001)&.002(.001)\\
1000& .001(.001)& .073(.012)&.008(.002)&.078(.007)& .035(.001)&.113(.002)\\
10000& .008(.002)&14.7(.474)&.826(.025)&15.9(.438)& 3.14(.083)&19.4(.263)\\ \hline \hline 
\end{tabular}
\caption{Mean computation time in seconds for Gini and Distance correlations with standard deviations in parentheses based on 100 repetitions. \label{tab:computation}}  
\end{table}
From Table \ref{tab:computation}, it is clear to see the computational advantages of the Gini correlation over the distance correlation.   Especially in $d=1$ with $n=10000$, computing Gini correlation takes 0.008 second, while it needs 14.7 seconds for the distance correlation. For $d>1$, the computation complexity of both correlations is $O(n^2)$ and the advantage of Gini correlation over the distance correlation is not as huge as that in $d=1$. The computation time of $\rho_g$ is about 19 times and 6 times as fast as that of $\rho_d$ for $d=10$ and $d=100$, respectively.  It is worthwhile to mention that we have made a R package called ``GiniDistance" available, in which the Rcpp version of Gini correlation function is 6 times faster than the current R code. 

Last, to illustrate robustness of the method with a small $\alpha$ value, we generate random variables from balanced 2-class Cauchy mixtures with Class 1 centered at 0 and Class 2 centered at $\delta$ changing from 0 to 1 . We take three values of  $\alpha = 0.5, 0.8,1$. Table \ref{tab:cauchy} reports the level and power of the permutation tests based on each $\alpha$ value. The sizes of three tests are similar to each other. As expected,  $\alpha=1$ performs inferior to the other two since the Cauchy distribution has no first moment. The test with $\alpha=0.5$ yields the highest power among the three. 
\begin{table}[H]
\center
\begin{tabular}{lc|ccccc} \hline\hline
$n$& $\alpha$& $\delta=0$ & $\delta=0.25$ &$\delta=0.5$ &$\delta=0.75$ &$\delta=1$ \\ \hline
60& 0.5 &0.059 &0.092& 0.171 &0.362 &0.565 \\ 
& 0.75 &0.064 &0.092 &0.139& 0.307& 0.475 \\ 
& 1 &0.055 &0.076& 0.120 &0.227 &0.382 \vspace*{0.2cm}\\  
120& 0.5 &0.070 &0.118 &0.334 &0.647& 0.876 \\ 
& 0.75 &0.069 &0.110 &0.269 &0.545 &0.791 \\ 
& 1 &0.070 &0.086 &0.199 &0.393& 0.633 \\ \hline \hline
\end{tabular}
\caption{Size and power of the Gini correlation permutation test based on different values of $\alpha$ under the Cauchy mixtures. \label{tab:cauchy}}  
\end{table}

\subsection{Real Data Application}

Three data sets from UCI Machine Learning Repository \cite{Dua17} are studied for dependence of categorical variable and numerical variables. 

The first data set is the famous Iris data set with the measurements in centimeters on sepal length and width and petal length and width, for 50 flowers from each of 3 species of iris.  Table \ref{tab:iris} lists Gini and Distance covariances/correlations between Species and each of measurements, also between Species and all measurements. Note that values in each column in Table \ref{tab:iris} are not comparable because they estimate difference quantities. Across each row, we can conclude that iris species have higher correlation with petal size than sepal size. Gini correlation estimators have smaller standard deviations than the distance correlation. Hence the Gini correlation estimators are more statistically efficient. Consequently, they lead to shorter confidence intervals than the distance correlation estimators.   
\begin{table}[h]
\caption{Correlations between Species and all variables,  between Species and each of variables for the Iris data.  Standard deviations are in parentheses.\label{tab:iris}}
\center
\begin{tabular}{l|ccccc} 
\hline\hline
      & All &   Sepal.L   &  Sepal.W    & Petal.L & Petal.W    \\ \hline
gCov& 0.794 (.041)& 0.376 (.043)& 0.109 (.022)& 1.530 (.076)& 0.654 (.034)\\
$\hat{\rho}_g$ &0.624 (.019)& 0.398 (.035)& 0.223 (.039)& 0.773 (.018)& 0.753 (.019)\\
dCov&0.529 (.050)& 0.125 (.018)& 0.036 (.008)& 0.510 (.047)& 0.218 (.019)\\
$\hat{\rho}_d$ &0.749 (.033)& 0.475 (.053)& 0.286 (.054)& 0.764 (.031)& 0.779 (.028) \\
\hline\hline
\end{tabular}
\end{table} 

The second data set is Letter Recognition Data Set of sample size 20000 on 16 features about 26 capital letters in the English alphabet. The black-and-white rectangular pixel character images were based on 20 different fonts and each letter within these 20 fonts was randomly distorted to produce a file of 20,000 unique stimuli. Each stimulus was converted into 16 primitive numerical attributes (statistical moments and edge counts) which were then scaled to fit into a range of integer values from 0 through 15.  Dependence covariance and correlation of each numerical feature to letter category are computed. Table \ref{tab:letter} reports the maximum and minimum covariances and correlations.   All methods identify the second feature (V3: the vertical position of box) has the weakest dependence with letter. The strongest dependence feature is the 11th feature (V12: the mean of $x^2y$) according to gcov, gcor and dcor.  Due to its large sample size, estimating the standard deviation of dcov or dcor is too time-consuming and hence is skipped. But for gcov and gcor, we are still able to estimate their standard deviations in a manageable computation time period.  This demonstrates a computational advantage of the Gini approach. 

\begin{table}[H]
\center
\begin{tabular}{l|llllr} 
\hline\hline 
& All & Max & Argmax & Min & Argmin \\ \hline
gcov & 0.638 (.0006) & 0.302 (.0005) & V12 & 0.0040 (.0006) & V3 \\
gcor& 0.201 (.0006)  & 0.418 (.0005) & V12 & 0.0023 (.0006) & V3 \\
dcov& 0.098 & 0.047 & V13 & 0.0003 & V3\\
dcor & 0.205 & 0.152 & V12 & 0.0008 & V3 \\
\hline\hline
\end{tabular}
\caption{Covariances and correlations between Letter and all variables.  The maximum and minimum covariance/correlation among letter and each of variables. Standard deviations are in parentheses.\label{tab:letter}}
\end{table}

\begin{figure}[h]
\centering
\includegraphics[width=12cm]{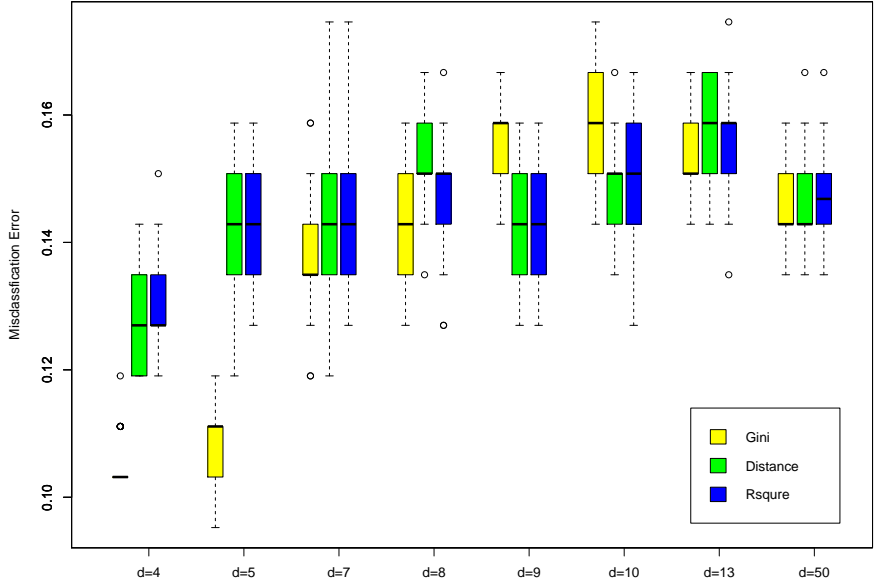}
\caption{Boxplot of out-of-bag classification errors of random forest classifier based on $d$ highest correlated features selected by Gini, distance and Pearson $R^2$ correlations.\label{fig:selection} } 
\end{figure}

The third data is LSVT Voice Rehabilitation Data Set about phonations evaluated as `acceptable' or `unacceptable' after speech rehabilitation treatments in Parkinson's disease based on 309 attributes on 126 samples.  Refer to Tsanas \cite{Tsanas14} for details of the data set and dysphonia measure attributes.  This data set has the dimension much larger than the sample size.  Our goal is to demonstrate a simple feature selection based on the highest correlated variables with the response class so that the selected subset of attributes is able to effectively predict a phonation as `acceptable' or `unacceptable'. We evaluate the selection method by its performance of the selected feature set in classification. We use Random Forest as the classifier due to its simplicity, popularity and effectiveness. Once $d$ features are selected, the random forest classifier (R package randomForest) with default parameters is applied and out of bag (oob) miss-classfication error is used as the performance criterion. As a benchmark, the oob error of the random forest classifier that uses all 309 predictors has a median of 0.167. The boxplots of oob errors based on 50 repetitions of random forest classifiers on each of $d$ values are provided in Figure \ref{fig:selection}.  The top one, two and three features selected by the Gini correlation coincide with those by the distance correlation. The feature 60 is the fourth highest correlated variable in terms of Gini correlation, while the feature 151 is the $4^{th}$ ranked according to the distance correlation.  Based on the top 4 features selected by Gini correlation, oob errors has a median of 0.103, significantly better than the median of 0.127 for the distance method and the $R^2$ method. It is worthwhile to mention that the variation of oob errors for  the model selected by Gini method is extremely small with a standard deviation of 0.003 and a median absolute deviation of 0, indicating stability of the selected model.    The attribute 80 ranks the fifth in three methods. However, its inclusion degrades performance. The error medians increase to 0.111, 0.143 and 0.143 respectively in three models. The model with top 6 correlated features selected by Gini and the distance methods is identical, and hence skipped in the boxplot. The differences of the top 7 and 8 features are the attribute 82 in the Gini method and the attribute 154 in the distance method. The Gini method yields better performance. However, when considering the top 9 and the top 10 features,  the distance and Pearson $R^2$ methods are better since they select the feature 155.  For $d=13$, the median error rate of the Gini selected model is about 0.8\% smaller than the one based on the other two methods. For $d=50$, Gini and distance methods produce a model with a similar performance in classification.   In general, the distance correlation and Pearson $R^2$ selection methods perform similarly and Gini feature selection performs better than the other two. The model with 4 features selected by Gini correlation has the best performance. 

\section{Conclusion and Future Work}
\label{sec:conclusion}

We have proposed the Gini correlation to measure dependence between categorical and numerical variables. The Gini correlation takes advantages of the nature of the categorical variable and hence has a simpler formulation than the distance correlation. As a result, the sample Gini correlation is more computationally and statistically efficient than the sample distance correlation.  Like Pearson $R^2$ correlation, Gini correlation has a nice interpretation as the ratio of the between variation and the total variation. Unlike Pearson $R^2$,  Gini correlation characterizes independence. This property is also possessed by the mutual information correlation (Gao et al. 2017), but it is difficult to estimate, especially for high-dimensional $\bi X$ in which density estimation and nearest neighborhood estimation suffer from the curse of dimensionality.  

Although the proposed Gini correlation has advantages over the existing correlations, it has some limitations. It is only orthogonal invariant but not affine invariant in general. One way to make it affine invariant is to consider the standardized samples $\bi z_i$ defined by  
$\bi z_i = S^{-1/2} \bi x_i $, where $S$ is the sample covariance matrix of ${\cal D} = \{\bi x_1,...,\bi x_n\}$. Then an affine Gini correlation estimator can be defined as 
$$ \hat{\rho}_G(\bi X,Y;\alpha) =\hat{\rho}_g(\bi Z, Y; \alpha)=  \frac{\hat{\Delta}(\bi Z; \alpha) -\sum_k\hat p_k \hat{\Delta}_k(\bi Z, \alpha)}{\hat{\Delta}(\bi Z; \alpha)}. $$
For the purpose of robustness, $S$ can be chosen to be some robust shape matrix estimator such as M-estimator, S-estimator (Shevlyakov and Oja 2016).   

Affine invariance property preserves an equivalent problem for statistical inference under linear transformations of data.  More desired property for a dependence measure is invariant under monotone transformations (Renyi 1959). We would like to have a dependence Gini measure such that 
$$\rho (\bi X, Y ) = \rho( g(\bi X), Y), $$
where $g$ is a one-to-one function. If $X$ is one-dimensional, one option shall be rank-based Gini correlation. Rather than the values of $X$, its ranks are used in calculation of Gini correlation. The rank-based approach preserves the monotonicity relationships and is robust against outliers. However, it may lose too much statistical efficiency. Continuities of the research in this direction are worthwhile. 

Another research direction is about the choice of parameter $\alpha$ in Gini distance correlation and distance correlation,  which corresponds to a distance metric in $\mathbb{R}^d$.    We can choose $\alpha$ so that the the association among variables is maximized.  The resulting correlation is called the maximal distance correlation (MDC). The reasoning to do so is that if a relationship exists, then the MDC is sensitive to capture it.  Studying the properties of MDC and developing efficient algorithms for the maximal distance correlations are focuses of research in the near future.

\clearpage
\pagenumbering{arabic}
\setcounter{page}{1}

\newpage
{\center{\Large \bf Online Supplemental File}}

\vspace{0.5cm} 
\noindent All necessary proofs of Remarks and derivations of Examples are provided in this file. 

\noindent {\bf Proof of Remark \ref{rem:2-3}}. For $d=1$ and $\alpha=1$, we would like to show that 
\begin{equation} \label{eqn:Fpsi}
T(X_k, X; 1) = \frac{1}{\pi} \int_{-\infty}^{\infty} \frac{|\psi_k(t)-\psi(t)|^2}{t^2} d t = 2 \int_{-\infty}^{\infty} (F_k(x)-F(x))^2 dx. 
\end{equation}
This is true because of the Parseval-Plancherel identity. Let the Fourier transforms of $F_k(x)$ and $F(x)$ are $\Psi_k(t)$ and $\Psi(t)$, respectively. By the Parseval-Plancheral formula,
\begin{equation} \label{eqn:PPf}
2 \pi \int_{-\infty}^{\infty} (F_k(x)-F(x))^2 dx  = \int_{-\infty}^{\infty} |\Psi_k(t)-\Psi(t)|^2 d t.
\end{equation}
Also taking derivatives of $\Psi(t)$ and $\Psi_k(t)$ with respect to $x$, we obtain that $\Psi(t) = -\psi(t)/(it)$ and $\Psi_k(t) = -\psi_k(t)/(it)$. Plugging them to (\ref{eqn:PPf}) proves  (\ref{eqn:Fpsi}).  

For $d>1$, although the Parseval-Plancheral identity still holds in $\mathbb{R}^d$, but there is no implicit relationship betweeen $\Psi$ and $\psi$. Hence, $T(\bi X_k, \bi X; 1)$ has no interpretation as the $L_2$ distance between $F_k$ and $F$.  \hfill$\square$

\noindent {\bf Proof of Remark \ref{rem:2-4}}.  It is sufficient to prove $\sum_{k=1}^Kp_k |\psi_k-\psi|^2 = \sum_{1\leq k\le l\leq K} p_k p_l |\psi_k-\psi_l|^2$. With the fact that $\psi = \sum_{l=1}^K p_l \psi_l$, we have 
\begin{align*}
& \sum_{k=1}^Kp_k |\psi_k-\psi |^2 = \sum_{k=1}^K p_k |\psi_k - \sum_{l=1}^Kp_l\psi_l|^2 =  \sum_{k=1}^K p_k |\sum_{l=1}^Kp_l (\psi_k-\psi_l)|^2 \\
& = \sum_{k=1}^K p_k\left [ \sum_{l=1}^K p_l^2|\psi_k-\psi_l|^2+ \sum_{l \neq m} p_lp_m (\psi_k -\psi_l)\overline{(\psi_k-\psi_m)} \right]\\
&= \sum_{1\neq k\le l\neq K}\left( p_k p_l |\psi_k -\psi_l|^2 (\sum_{m=1}^K p_m)\right) 
\end{align*}
The last equation is due to combining the terms $p_kp_lp_m (\psi_k-\psi_l)\overline{(\psi_k-\psi_m)}$ and $p_kp_lp_m (\psi_k-\psi_l)\overline{(\psi_m-\psi_l)}$ together. Since $\sum p_m =1$, the remark is proved.  \hfill{$\square$}

\noindent {\bf Proof of Remark \ref{identity}}. 
On one hand, 
\begin{align*}
& \E |\bi X-\bi X^\prime|^\alpha|Y-Y^\prime|^\alpha+\E |\bi X-\bi X^\prime|^\alpha\E|Y-Y^\prime|^\alpha-2\E |\bi X-\bi X^\prime|^\alpha|Y-Y^{\prime\prime}|^\alpha \\
& = \sum_{k\neq l} p_kp_l \Delta_{kl}(\alpha) + \Delta(\alpha) (1-\sum_j p_j^2) -2 \sum_k p_k(1-p_k) \E|\bi X-\bi X_k|^\alpha \\
&= \sum_{k \neq l} p_kp_l \Delta_{kl}(\alpha) + \sum_{k \neq l} p_kp_l \Delta_{kl}(\alpha)(1- \sum_j p_j^2) + \sum_{k}p_k^2 \Delta_k(\alpha) ((1- \sum_j p_j^2) \\
&- 2 \sum_k p_k(1-p_k)(p_k\Delta_k(\alpha)+\sum_{k \neq l} p_l \Delta_{kl}(\alpha)) \\
&= \sum_{k \neq l} p_k p_l (2 p_k -\sum p_j^2) \Delta_{kl}(\alpha)+\sum _k p_k^2(2p_k -1 -\sum_j p_j^2) \Delta_k(\alpha).
\end{align*}
On the other hand, by \eqref{charac}
\begin{eqnarray*}
dCov(\bi X,Y; \alpha) &=& \sum_k p_k^2 T(\bi X_k, \bi X; \alpha)\\
&=&\sum_k p_k^2 (2 \E |\bi X_k - \bi X|^\alpha - \E |\bi X_k-\bi X_k^\prime|^\alpha -\E |\bi X-\bi X^\prime|^\alpha) \\
&=& \sum_k p_k^2[2 ( \sum_{l\neq k} p_l \Delta_{kl} (\alpha)+ p_k \Delta_k(\alpha))] -\sum_k p_k^2 \Delta_k(\alpha) -(\sum_k p_k^2) \Delta(\alpha) \\
& =& \sum_{k\neq l} (2p_k - \sum_j p_j^2)\Delta_{kl}(\alpha) + \sum _k p_k^2(2p_k -1 -\sum_j p_j^2) \Delta_k(\alpha).
\end{eqnarray*}
The result of Remark \ref{identity} is proved.  
\hfill{$\square$}

\noindent {\bf Proof of Equation (\ref{dcovyy})}.  We have 
\begin{eqnarray*}
dCov(Y,Y)  &=&  \E |Y-Y^\prime|^{2 } + (\E |Y-Y^\prime|)^2 - 2 \E|Y-Y^\prime| |Y-Y^{\prime\prime}| \\
&=& \sum _k p_k(1-p_k) + (\sum _k p_k(1-p_k))^2 -2 \sum _kp_k(1-p_k)^2\\
&=& 1- \sum_k p_k^2 +(1-\sum _k p_k^2)^2- 2+4\sum_k p_k^2 -2 \sum_k p_k^3\\
&=& \sum p_k^2 + (\sum_k p_k^2)^2 -2 \sum_k p_k^3.
\end{eqnarray*}
\hfill{$\square$}

\noindent {\bf Proof of Remark \ref{rem:2class}}. For $K=2$  with $p_1+p_2=1 $,  we have
 \begin{align*}
&dCov(Y,Y) = p_1^2+p_2^2+(p_1^2+p_2^2)^2-2(p_1^3+p_2^3)= 4p_1^2p_2^2; \\
&p_1^2 |\psi_1-\psi|^2+p_2^2|\psi_2-\psi|^2=p_1^2 |\psi_1-p_1\psi_1-p_2\psi_2|^2+p_2^2 |\psi_2-p_1\psi_1-p_2\psi_2|^2=2p_1^2p_2^2 |\psi_1-\psi_2|^2; \\
&p_1|\psi_1-\psi|^2+p_2|\psi_2-\psi|^2= p_1p_2^2|\psi_1-\psi_2|^2+ p_1^2p_2 |\psi_1-\psi_2|^2 = p_1p_2 |\psi_1-\psi_2|^2.
\end{align*}
Together with the definitions of  $gCov(\bi X, Y;\alpha)$  and $dCov(\bi X, Y;\alpha)$, 
Remark \ref{rem:2class} is proved. \hfill{$\square$}\\

\noindent {\bf Detailed Derivations for Examples}.
We use the Theorem 3.2 result of Edelmann, Richards and Vogel (2017), which states that
\begin{equation}\label{eqn:dcovxx}
dCov(X,X) = 8 \int\int_{-\infty <x<z<\infty} F^2(x)(1-F(z))^2 dz dx. 
\end{equation}
Due to the difficulty to evaluate  (\ref{eqn:dcovxx}), we only provide the distance correlation formula in Example 1 where $X \sim p Exp(\theta)+(1-p) Exp(\beta)$. 

\noindent {\bf Example 1}.   $ X_1 \sim Exp(\theta)$ and $X_2  \sim Exp(\beta)$ are independent. Then,
\begin{align*}
\Delta_{12}= \E |X_1-X_2| = \int_0^\infty\int_0^\infty|x_1-x_2| \frac{1}{\theta} e^{-x_1/\theta}\frac{1}{\beta}e^{-x_2/\beta} d x_1 d x_2 = \frac{\theta^2+\beta^2}{\theta+\beta}.
\end{align*}
Plugging $F(x) = p(1-e^{-x/\theta})+(1-p)(1-e^{-x/\beta}) = 1-pe^{-x/\theta}-(1-p)e^{-x/\beta}$ in (\ref{eqn:dcovxx}) and following a tedious evaluation of the integral, we have the result of $dCov(X,X)$. 

To prove that the squared Pearson correlation and Gini correlation are increasing with $r>1$, we obtain their derivatives with respect to $r$ as follows. 
\begin{align*}
& \frac{\partial \rho_g}{\partial r} = \frac{p(1-p)(r-1)[ 4p-p^2+(1-2p+2p^2)(r-1)]}{(p+(1-p)r^2+p(1-p)(1-r)^2)^2} >0\\
& \frac{\partial \rho_p^2}{\partial r} = \frac{2p(1-p)(r-1)[p+(1-p)r^2+p(1-p)(1-r)^2+(1-p)(r-1)((1+p)r-p)]}{(p+(1-p)r^2+p(1-p)(1-r)^2)^2}>0
\end{align*}
The positiveness of those derivatives proves the claim.  \hfill$\square$ 

\noindent {\bf Example 2}. $ X_1 \sim N(\mu_1, \sigma^2$ and $X_2  \sim N(\mu_2, \sigma^2)$ are independent. Let $a = |\mu_1-\mu_2|/\sigma$. Then, $Z= \frac{X_1-X_2}{\sqrt {2}\sigma} \sim N(a/\sqrt{2},1)$ and 
\begin{align}
&\Delta_{12}= \E |X_1-X_2| =\sqrt{2}\sigma \E |Z| = \sqrt{2}\sigma \left(\sqrt{2} a \Phi(\frac{a}{\sqrt{2}})+2 \phi (\frac{a}{\sqrt{2}})-\frac{a}{\sqrt{2}}\right) \nonumber \\
&= \sigma\left( 2a \Phi( \frac{a}{\sqrt{2}})+2\sqrt{2} \phi (\frac{a}{\sqrt{2}})-a\right),\label{delta12}
\end{align}
where $\Phi$ and $\phi$ are the cumulative distribution function and probability density distribution function of $N(0,1)$, respectively. 
Also $\Delta_1 = \Delta_2 = 2\sigma/\sqrt{\pi}$ is obtained by (\ref{delta12}) with $a =0$. 

To prove the monotonecity property of $\rho_g$ in $a$ for any $p$,    it is sufficient to prove that $g(a) : = 2 a \Phi(a/\sqrt{2} )+2\sqrt{2}\phi(a/\sqrt{2})-a$ is increasing in $a$. This can be done by showing that 
$$\frac{\partial g(a)}{\partial a} = 2 \Phi(a/\sqrt{2}) -1>0.$$ 

To obtain the maximum correlations with respect to $p$, we have 
\begin{align*}
&\frac{\partial \rho_g}{\partial p} = \frac{(1-2p)(g(a)-2/\sqrt{\pi})}{[(p^2+(1-p)^2+\sqrt{\pi}p(1-p)g(a)]^2}:=0
\end{align*}
With an additional check of $\frac{\partial^2 \rho_g}{\partial^2 p} \vert_{p=0.5}<0$, we conclude that the maximum Gini correlation is achieved at $p=0.5$.  \hfill$\square$ 

\noindent {\bf Example 3}. $X_1 \sim N(\mu, \sigma_1^2)$ and $X_2  \sim N(\mu_2, \sigma_2^2)$ are independent. Then, $X_1-X_2 \sim N(0, \sigma_1^2+\sigma_2^2)$. By (\ref{delta12}) with $a=0$ and $\sqrt{2}\sigma=\sqrt{\sigma_1^2+\sigma_2^2}$,   we have $\Delta_{12}=  \sqrt{2(\sigma_1^2+\sigma_2^2)}/\sqrt{\pi}$.

The first derivatives of the correlations with respective to $r$ are 
\begin{align*}
& \frac{\partial \rho_g}{\partial r} = \frac{2p(1-p)(r-1)[2p(2-p) +(1-2p+2p^2)(r-1)]}{[(2p-p)^2+(1-p^2)r^2+(1-2p+2p^2)r]^2} >0\\
& \frac{\partial \rho_p^2}{\partial r} = \frac{2p(1-p)(r-1)[p+(1-p)r]}{[p+(1-p)r^2+p(1-p)(1-r)^2]^2}>0.
\end{align*}
This completes the claim that the correlations are increasing in $r>1$.  \hfill$\square$

\end{document}